\documentclass[sigconf]{acmart}
\usepackage{enumitem}
\usepackage{soul}
\usepackage{multirow}
\usepackage{soul}
\usepackage{balance}

\newcommand{\cmt}[1]{\ignorespaces}

\newcommand{\sys}{StoryBuddy}

\newcommand{\yxhighlight}[1]{{\textcolor{black}{#1}}}

\newcommand{\tlhighlight}[1]{{\textcolor{black}{#1}}}
\newcommand{\zzhighlight}[1]{{\textcolor{black}{#1}}}

\AtBeginDocument{%
  \providecommand\BibTeX{{%
    \normalfont B\kern-0.5em{\scshape i\kern-0.25em b}\kern-0.8em\TeX}}}

\copyrightyear{2022}
\acmYear{2022}
\setcopyright{acmlicensed}\acmConference[CHI '22]{CHI Conference on Human
Factors in Computing Systems}{April 29-May 5, 2022}{New Orleans, LA, USA}
\acmBooktitle{CHI Conference on Human Factors in Computing Systems (CHI
'22), April 29-May 5, 2022, New Orleans, LA, USA}
\acmPrice{15.00}
\acmDOI{10.1145/3491102.3517479}
\acmISBN{978-1-4503-9157-3/22/04}

\copyrightyear{2022}
\acmYear{2022}
\setcopyright{acmlicensed}\acmConference[CHI '22]{CHI Conference on Human Factors in Computing Systems}{April 29-May 5, 2022}{New Orleans, LA, USA}
\acmBooktitle{CHI Conference on Human Factors in Computing Systems (CHI '22), April 29-May 5, 2022, New Orleans, LA, USA}
\acmPrice{15.00}
\acmDOI{10.1145/3491102.3517479}
\acmISBN{978-1-4503-9157-3/22/04}

\begin{document}

\title[StoryBuddy: A Human-AI Collaborative Agent for Parent-Child Interactive Storytelling]{StoryBuddy: A Human-AI Collaborative Chatbot for Parent-Child Interactive Storytelling with Flexible Parental Involvement}


\author{Zheng Zhang}
\authornote{equal contribution}
\email{zzhang37@nd.edu}
\affiliation{%
  \institution{University of Notre Dame}
   \city{Notre Dame}
   \state{IN}
  \country{USA}
}
\author{Ying Xu}
\authornotemark[1]
 \email{ying.xu@uci.edu}
\affiliation{%
  \institution{University of California, Irvine}
    \city{Irvine}
    \state{CA}
    \country{USA}
}
\author{Yanhao Wang}
 \email{yanhaow@gatech.edu}
\affiliation{%
  \institution{Georgia Institute of Technology}
    \city{Atlanta}
    \state{GA}
    \country{USA}
}
\author{Bingsheng Yao}
\email{yaob@rpi.edu}
\affiliation{%
  \institution{Rensselaer Polytechnic Institute}
    \city{Troy}
    \state{NY}
    \country{USA}
}
\author{Daniel Ritchie}
 \email{drritchi@uci.edu}
\affiliation{%
  \institution{University of California, Irvine}
    \city{Irvine}
    \state{CA}
    \country{USA}
}
\author{Tongshuang Wu}
 \email{wtshuang@cs.washington.edu}
\affiliation{%
  \institution{University of Washington}
    \city{Seattle}
    \state{WA}
    \country{USA}
}
\author{Mo Yu}
 \email{yum@us.ibm.com}
\affiliation{%
  \institution{IBM Research}
    \city{Yorktown Heights}
    \state{NY}
    \country{USA}
}
\author{Dakuo Wang}
\authornote{corresponding authors}
 \email{dakuo.wang@ibm.com}
\affiliation{%
  \institution{IBM Research}
    \city{Cambridge}
    \state{MA}
    \country{USA}
}
\author{Toby Jia-Jun Li}
\authornotemark[2]
 \email{toby.j.li@nd.edu}
\affiliation{%
  \institution{University of Notre Dame}
   \city{Notre Dame}
   \state{IN}
  \country{USA}
}

\renewcommand{\shortauthors}{Zhang et al.}

\begin{abstract}

Despite its benefits for children's skill development and parent-child bonding, many parents do not often engage in interactive storytelling by having story-related dialogues with their child due to limited availability or challenges in coming up with appropriate questions. While recent advances made AI generation of questions from stories possible, the fully-automated approach excludes parent involvement, disregards educational goals, and underoptimizes for child engagement. Informed by need-finding interviews and participatory design (PD) results, we developed StoryBuddy, an AI-enabled system for parents to create interactive storytelling experiences. \tlhighlight{StoryBuddy's design highlighted the need for accommodating dynamic user needs between the desire for parent involvement and parent-child bonding and the goal of minimizing parent intervention when busy. The PD revealed varied assessment and educational goals of parents, which StoryBuddy addressed by supporting configuring question types and tracking child progress. A user study validated StoryBuddy's usability and suggested design insights for future parent-AI collaboration systems.}



\end{abstract}




\begin{teaserfigure}
        \includegraphics[width=\linewidth]{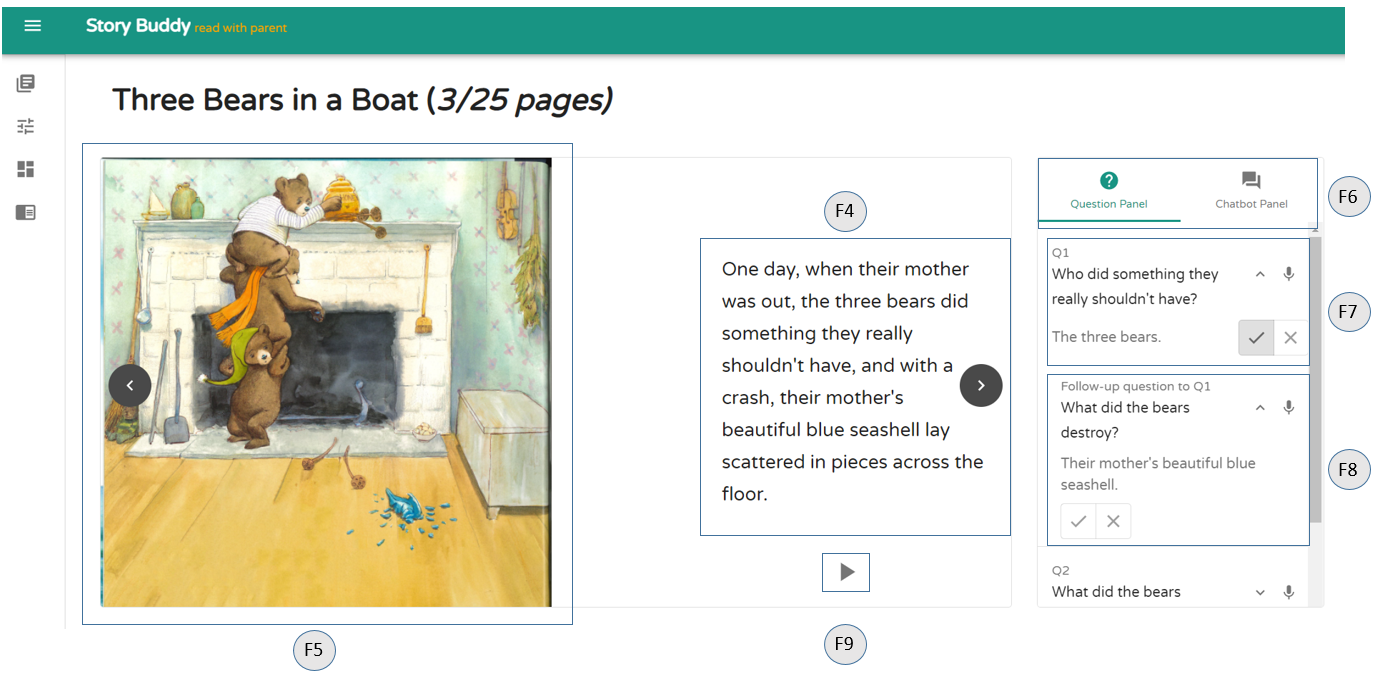}
        \caption{StoryBuddy's story reading interface in the parent-AI co-reading mode}
        \label{fig:parentReading}
    \end{teaserfigure}

\maketitle

\section{Introduction}

    Storytelling is a common parent-child activity that provides many educational benefits such as improving children's language fluency, communication skills, cultural and emotional awareness, and other aspects of cognitive development~\cite{peck1989using, wright1995storytelling}. Interactive storytelling in particular, where a storyteller asks questions relevant to story content and prompts a child to express their thoughts about the story, has been shown to maximize the educational benefits of storytelling \cite{kotaman2020impacts}. Many parents experience barriers like difficulty in coming up with appropriate questions, high cognitive load from multi-tasking, and challenges with keeping track of the child's progress. To address these barriers, many digital systems, both from the industry and research community, have been proposed to facilitate parent-child interactive storytelling with children.
    

    
    Many prior digital interactive storytelling systems have been shown to be effective in supporting various learning goals. For example, StoryCoder~\cite{dietz_storycoder_2021} leverages storytelling as a creative activity by allowing children to first listen to stories and then modify these stories in computational thinking games. This approach was shown to be effective in the development of computational thinking. Conversational agents were also developed to support children's literacy development ~\cite{xu_current:_2021}, bilingual language acquisition~\cite{bhatti2021conversational}, and foster science learning~\cite{xu2020using}. In the HCI community, several empirical studies have been done to investigate how children and parents interact with existing voice agents such as Amazon Alexa and Google Home~\cite{beneteau_parenting:_2020, xu2020exploring, xu2020you, wiederhold2018alexa, lovato2019hey}. These studies identified opportunities in the use of voice agents to facilitate learning, development, and social goals of children, but also pointed out challenges in facilitating child-agent interaction. Commercial products such as Luka~\cite{ling_luka_2021} and Codi~\cite{pillar_learning_meet_2021} are AI-enabled robot toys that can facilitate interactive storytelling experiences. Codi is a storytelling robot that can tell over 100 pre-recorded stories. Luka is a ``AI reading companion'' that the child can place in front of a book while the child reads the book. Luka can recognize the book (from ~20,000 books in the developer's library) and ask the child preset questions relevant to the story. An important limitation of all these existing systems for interactive storytelling is that their questions are manually crafted---therefore they only support a limited set of books or stories that the system developers prepared.

    
    Recent advances in natural language comprehension and question generation (e.g.,~\cite{du-etal-2017-learning, shakeri-etal-2020-end, labutov-etal-2015-deep}) made it feasible to automatically generate question-answer pairs (QA pairs) about story plots from \textit{any} storybooks, enabling fully automated interactive question-answering between children and a chatbot. But there are several issues in the adoption of this approach in real-life storytelling sessions with children: 
    \begin{enumerate}[topsep=0pt]
        \item While the vast majority of generated QA pairs are syntactically and factually correct, many do not serve educational purposes (e.g., too trivial, not relevant to the main story plot) and are not necessarily appropriate (e.g., containing difficult words) for the children~\cite{hill2015goldilocks, mostow_generationg:_2009}.
        
        \item A fully automated approach excludes parent involvement---prior research shows that parent storytelling not only develops language and comprehension skills of the children but also strengthens the bond between parents and children~\cite{frude2011family, vasalou2020designing}.
        
        \item Simply asking questions in sequence from a list of generated questions does not optimize for child engagement due to the lack of logical connection between questions and the lack of guidance for the children in case of confusion or incorrect answers.  
    \end{enumerate}


    To address the above limitations of an \textit{AI-only} approach, we explore a \textit{human-AI collaboration} approach that incorporates the expertise and preferences of parents into the development of interactive storytelling experiences for their children. However, there is no one-size-fits-all solution. Our formative study and participatory design process (Sections~\ref{sec:formative_study} and~\ref{sec:design_process}) found that parents have different motivations, objectives, and preferences for using digital storytelling systems, and they want a system that can adapt to various usage scenarios. This large variety of user needs results in diverse, sometimes even conflicting, design goals and constraints for the system. For example, parents reported that they regard storytelling as an important way to strengthen relationships between themselves and their children, therefore it is important for any AI assistance in storytelling to preserve direct parent-child interactions. At the same time, they expressed the desire for an automated storytelling system that can keep their children engaged without any parent intervention in situations where they need to focus on something else (e.g., when they are in a meeting while working from home). We also heard different opinions from parents on whether they prefer to put a stronger emphasis on the skill development and assessment objectives in storytelling or if they wish to just ``keep it fun'' for their children as a form of entertainment.  
    


    
    Building on prior literature and our own formative investigation with 10 families, we designed and developed StoryBuddy, a new system that allows parents to collaborate with AI in creating storytelling experiences with interactive questioning-answering. Through co-design sessions with four parents using storyboards, we proposed an interaction strategy that supports two distinct modes: (1) an assisted \textit{parent-AI co-reading mode} where the AI assists the parent in storytelling by identifying potential opportunities for asking questions and recommending follow-up questions. \sys{} in this mode can reduce the cognitive load and lower the literacy barrier for the parent, facilitate skill development for the child while encouraging direct parent-child interaction that both parties value in their relationship. (2) an asynchronous \textit{automated bot-reading mode} where the parent can create an interactive storytelling bot for \textit{any} story by configuring the question generation model, selecting from the generated questions, and customizing follow-up questions. The bot can then tell stories, ask children questions and provide feedback, and converse with the children to keep them engaged without intervention from the parents. \sys{} in both modes also tracks the child's progress and visualizes children's performance data in a dashboard, enabling the parent to assess the development of the child's comprehension skills.

    
    This paper presents the following three main contributions:
    \begin{itemize}[topsep=0pt]
        \item a formative interview study and a participatory design process with parents that uncover the large variations in parents' objectives for interactive storytelling, their need for the support of flexible parent involvement, their challenges with the high cognitive load from multitasking, and their desired strategies for enhancing child engagement in interactive storytelling.
        \item the design and implementation of \sys{}, a system where parents collaborate with AI in creating interactive storytelling experiences with question-answering for their children
        \item a user study with 12 pairs of parents and children that evaluates the usability of \sys{} and sheds light on how parents and children interact with \sys{}
    \end{itemize}
    
    \tlhighlight{From the findings of the design process, the implementation of the system artifact, and a user study with the system, this paper presents several implications for designing human-AI collaborative systems in facilitating parent-children interaction. Specifically, we (1) identified challenges in designing a workflow that accommodates effective \textit{partial} automation in real-time that copes with interruptions and resumptions, (2) presented a design strategy of transforming synchronous involvement into asynchronous involvement to support flexibility in parent involvement, and (3) discussed opportunities in designing flexible multi-faceted roles of an AI companion that fits into the existing parent-child interaction dynamic in a familiar activity (storytelling) while balancing between multiple educational, developmental, assessment, engagement, and relationship-building goals.}

\section{Related Work}
    \subsection{Digital Systems for Facilitating Interactive Storytelling}
    
    Storytelling with children is an activity that provides significant benefits in skill development, relationship building, and entertainment~\cite{peck1989using, wright1995storytelling, frude2011family}. Specifically, storytelling between parents and children is a routine activity in families across different cultures~\cite{shanahan2010national}. Interactive storytelling is a form of storytelling where, in addition to merely narrating the story verbatim, parents actively interact with their children about the story content. An effective and popular strategy in interactive storytelling is guided conversation~\cite{zevenbergen2003dialogic} (also known as dialogic reading), where a storyteller asks a child questions about story content and provides responsive feedback. This strategy allows children to actively participate in the storytelling process, reflect on their comprehension of the story, and express their understanding through multi-turn dialogues. Prior studies in this area found that guided conversation with question-answering has positive impacts on the development of language and literacy skills for children ~\cite{mol2008added, flack2018effects}.
    
    \yxhighlight{While there is evidence on the benefits of dialogic reading for children across a broad age range, much of the research attention has been focused on younger children aged three to eight who are at the stage of ``learning to read'' and do not have fluent decoding skills ~\cite{lever2011discussing, mol2008added}. Dialogic reading is particularly suitable for this age group as this reading activity is typically carried out orally, thus allowing young children to fully allocate their cognitive resources on making sense of the text they hear ~\cite{kim2017simple}. Thereby, dialogic reading  promotes the kind of oral language skills, including vocabulary and narrative comprehension~\cite{hargrave2000book}, that are strongly linked to children's later reading skills and academic success as they move to the ``reading to learn'' stage after the age of eight~\cite{storch2002oral}. Nevertheless, children aged three to eight span two different stages of the reading development ~\cite{norman1987stages, hoien1988stages}: a \textit{pre-reading stage} when children gain mastery over the sound structure of spoken language, make inferences of stories from pictures, and develop listening comprehension skills; and an\textit{ early-reading stage} when children learn to decode print text and begin to read fluently and strategically. Due to pre-reading children's limited decoding skills, they primarily rely on stories being read to them, while children in the early-reading stage start to read stories more independently. Our system provides audio narration to support pre-reading children, but also allows early-reading children to read without audio narration. We will discuss the different usage patterns between these two stages in the findings of our user study (Section~\ref{sec:obeservation_result}) and how future versions of StoryBuddy can better accommodate the specific needs of children in these age groups (Section~\ref{sec:future_work}).}

    \yxhighlight{Prior research has proposed several effective questioning strategies for dialogic reading. In general, there is a consensus that open-ended ``Wh-'' questions are more effective in eliciting children's verbal responses than yes-or-no or multiple-choice questions ~\cite{paris2003assessing}. Furthermore, ``Wh-''questions can be categorized based on the information required for formulating answers. Rubegni and colleagues suggested incorporating two types of ``wh-''questions: basic prompts that focus on children's \textit{recall} of story events and contexts, and ``Theory of Mind'' prompts that encourage children to \textit{make inferences} of story characters' thoughts, feelings, and intentions ~\cite{rubegni2021girl}. Blewitt's study also observed the benefits of including both types of prompts ~\cite{blewitt2009shared}. Thus, in our research, we follow these evidence-based suggestions to incorporate both recall and inferential questions in our question-generation model.} 
    
    Various digital tools have been introduced to help facilitate different aspects of interactive storytelling~\cite{xu_same:_2021}. For example, Kory and Breazeal created an embodied learning companion robot that can introduce new vocabulary words during storytelling~\cite{kory2014storytelling}. Michaelis and Mutlu designed an in-home learning companion robot that can make preprogrammed comments in stories~\cite{michaelis_2017_development}. StoryCoder presents two storytelling games and four computational thinking games to support the use of interactive storytelling as a way to teach computational thinking concepts~\cite{dietz_storycoder_2021}. There were also tools for supporting the creation of multimedia stories such as Fiabot~\cite{rubegni_2014_fiabot} and StoryBank~\cite{frohlich_2019_storybank}. Besides academic research work, commercial products such as Luka~\cite{ling_luka_2021} and Codi~\cite{pillar_learning_meet_2021} also support automated telling of stories and child-bot conversation about the story content.
    
    Compared with prior work, the two key novel contributions of StoryBuddy are (1) its support for interactive question-answering on \textit{any} stories; and (2) its new design features for supporting flexible levels of parent involvement. The systems we discussed above rely on manually prepared story-specific questions. In comparison, enabled by a state-of-art question-answer generation model, StoryBuddy can automatically generate appropriate questions, identify follow-up questions, and engage in multi-turn question-answering with children for any stories. Several parent-AI collaborative mechanisms in StoryBuddy help parents ensure that the generated questions (1) are appropriate for their child; and (2) can serve the intended goals parents have. Previous systems also lack support for flexible parent involvement, which is a key user need according to both prior literature~\cite{lin_parental:_2021} and our formative study findings. StoryBuddy's two distinct modes support situations for both when the parent is present and when the parent is absent. In StoryBuddy, parents who wish to have more control have the option to customize the question-answering content, select generated question types, and track child progress. For others, these steps can be automated with little parental intervention.


    \subsection{Studies of Child-Agent Interaction}
    \yxhighlight{As we discussed before, dialogic reading resolves around back-and-forth conversation between adults and children. This makes conversational agents favorably positioned to act as children's reading partners. This technology has the affordances to understand unconstrained natural language input, thus allowing for complex dialogue and potentially mimicking human-to-human spoken conversation. Researchers recognized that conversational technologies can potentially offer a potent new mechanism for teaching, engaging, and supporting children in daily life ~\cite{garg2020conversational}. The resulting developments may be especially valuable for young children, as their lack of proficiency in reading and writing cause difficulty for them to navigate many digital contents. } 
    
    The design of StoryBuddy was informed by not only the results from our formative study and participatory design process but also insights from prior studies on how children interact with conversational agents.
    
    Prior work identified opportunities in the use of conversational bots for facilitating learning, development, and social goals of children~\cite{beneteau_parenting:_2020, garg_conversational:_2020, mack_codesigning_2019, xu_2019_young, du_alexa:_2021,xu_current:_2021, xu_exploring_2020}. For example, a study by Beneteau et al. investigated how parents and children interact with Amazon Alexa in family homes through a 4-week deployment study. The study results suggested that the use of voice interfaces naturally promotes verbal communication and expands the communication skills of children. Parents found opportunities to use conversational bots to augment parenting practices. Specifically, they can complement parenting tasks (including storytelling) and increase the autonomy of their children~\cite{beneteau_parenting:_2020}. A co-design study by Garg and Sengupta suggested that a conversational agent can be an ideal learning companion for children. Especially, parents want these agents to include them in the learning activities and to allow them to monitor their children's use~\cite{garg_conversational:_2020}. Voice interfaces were also found to be effective in keeping children engaged~\cite{pantoja_voice_2019}. Findings from these studies motivated StoryBuddy's design strategies in child skill development and assessment through conversational question-answering, enhancing parental involvement and customizability, and supporting a dynamic interaction paradigm that combines parent-child interaction with child-agent interaction to improve child engagement.

    Prior studies also identified challenges specifically in facilitating child-agent interaction. While communication breakdown is common in general human-agent conversation~\cite{bentley2018understanding, Cowan:2017:IHY:3098279.3098539, Luger2016,li_sugilite:_2017}, and mechanisms such as~\cite{li_sovite:_2020, ashktorab2019resilient, myers_2018_patterns} have been proposed to handle them, children’s limited communication skills make it more difficult to avoid or repair communication breakdowns in child-agent interaction---Children are likely to encounter difficulties in understanding instructions, fail to follow the conversation flow, and struggle with appropriate turn-taking when interacting with agents~\cite{du_alexa:_2021,pantoja_voice_2019,lovato_2019_hey}. In StoryBuddy, parents are involved in the child-agent interaction in the parent-AI co-reading mode, which alleviates these challenges by allowing the parent to help with breakdown repair. In both modes of StoryBuddy, controls in the graphical user interface (GUI) are available alongside the voice interface, so that the parent or child can still proceed through multi-modal interaction~\cite{oviatt_ten_1999} in conversation breakdown situations~\cite{li_interactive:_2020}.

    \subsection{Systems for Question-Answer Generation}
    StoryBuddy belongs to a category of systems that automatically generate questions and answers for a given piece of text (known as QAG systems in the natural language processing (NLP) community), but StoryBuddy's design goals and intended context of use are quite different from the vast majority of existing QAG systems. Most works on QAG systems approach the problem from a pure machine learning perspective, trying to invent new rule-based (e.g.,~\cite{yao2012semantics, labutov-etal-2015-deep}) or neural-network-based (e.g.,~\cite{du-etal-2017-learning,scialom-etal-2019-self,dong2019unified, tang2017question, wang2017joint}) models that can generate ``more accurate'' questions and answers. This ``accuracy'' is commonly measured using objective similarity-based metrics (e.g., BLEU~\cite{papineni_bleu_2002} which measures the precision of \textit{n}-grams, ROGUE~\cite{lin-2004-rouge} which measures the recall of \textit{n}-grams) that compare the generated questions against the gold standard of human-generated questions. While these systems perform well in generating \textit{correct} and \textit{relevant} questions. Their generated questions usually lack educational values and are ineffective in maintaining child engagement, because these QAG systems are not optimized for these objectives. 
    
    Unlike prior systems, the question-answer generation model used in StoryBuddy was specifically designed to be ``as if a teacher or parent is to think of a question to improve children’s language comprehension ability while reading a story to them~\cite{yao2021ais}.'' and was trained on a dataset of children's storybooks annotated by educational experts for supporting interactive storytelling. Another category of relevant work is on interactive question-answering systems that seek to retrieve answers to questions that users ask (e.g.,~\cite{sun_towards_2006, rebanal_2021_xalgo}). These systems focus on answer retrieval (instead of question generation) and therefore have quite different goals from our work. \looseness=-1 
    
    Compared with prior work, StoryBuddy emphasizes interaction design and human-AI collaboration aspects. In StoryBuddy, parents are heavily involved in the pre-configuration, question selection, and follow-up question generation process to better adapt AI-generated questions to interactive storytelling. In comparison, most existing work in QAG only focuses on the model without considering the intended context of use and the goals of the users.  
    
\section{Formative Study}
    \label{sec:formative_study}
    As discussed above, there is a range of systems, tools, and applications aiming to support storytelling in early childhood. However, these solutions have not been designed to promote parent involvement or provide personalized reading experiences for individual families. This is less ideal for supporting children and families' diverse needs. To further understand this issue from the users' perspectives, we conducted a formative study to gather information on (1) families' daily practices of storytelling and digital device usage and (2) families' general needs and expectations of digital storytelling systems. 
    
\begin{table*}[]
 \def\arraystretch{1.1}
    \centering
\begin{tabular}{|c|c|c|c|c|c|c|}
\hline
\textbf{ID} & \textbf{\begin{tabular}[c]{@{}c@{}}Parent Gender\end{tabular}} & \textbf{\begin{tabular}[c]{@{}c@{}}Child Age\end{tabular}} & \textbf{\begin{tabular}[c]{@{}c@{}}Parent-Child\\ Language Use\end{tabular}} & \textbf{\begin{tabular}[c]{@{}c@{}}Minutes Reading\\ (Weekday)\end{tabular}} & \textbf{\begin{tabular}[c]{@{}c@{}}Minutes Reading\\ (Weekend)\end{tabular}} \\ \hline
BF1 & Female & 4 & English and Spanish  & 30 & 0 \\
BF2 & Female & 4 & Mostly Spanish & 30 & 10 \\
BF3 & Female & 4 & English and Spanish  & 5--10 & 0 \\
BF4 & Female & 6 & Mostly English  & 10--15 & 20 \\
BF5 & Female & 5 & Only Spanish  & 10 & 10--15 \\
BF6 & Female & 4 & Mostly Spanish & 90 & 30 \\
BF7 & Female & 7 & Mostly Spanish & 30 & 20 \\
BF8 & Female & 5 & Mostly English  & 30 & 0 \\
BF9 & Female & 4 & English and Spanish  & 20 & 0 \\
BF10 & Female & 5 & English and Spanish  & 20 & 0 \\ \hline
\end{tabular}
\caption{Demographics of formative study participants and the average daily time they spent on storybook reading.}
\label{tab:interviewee_demo}
\end{table*}
    
    \subsection{Method}
    We recruited and interviewed ten families with at least one child aged three to eight years from two different communities in the Western U.S., including one predominately White and Asian University community and one nearby working-class, Spanish-speaking community. \yxhighlight{Detailed participant information is displayed in Table 1. As shown in the table, all participants on average spent sometime on storybook reading daily. The ``Parent-Child Language Use'' column indicates the language used at home between parents and children.} Each interview session followed a semi-structured format, in which we asked parents questions about how they used digital media and devices for storybook reading with their children. We also asked parents about their general attitudes toward existing storytelling applications and their suggestions for improving these applications. We purposely started the interview with broad questions that placed fewer restrictions on the participants' responses and then asked more focused follow-up questions to probe parents' elaborations on certain topics. The interviews lasted 60 minutes for each participant and were carried out via video conferencing.
    
    \yxhighlight{We used an inductive process to analyze the interviews. We began with qualitative memoing ~\cite{birks2008memoing}, in which members of the research team viewed the same portion of the data together, with each researcher individually memoing their own notes. After specific intervals (usually 5 minutes) researchers would pause data playback and discuss with one another the meaning that emerged from the data. During this process, we noticed emerging themes related to parents' perception and expectation of digital reading technologies. We then systematically coded all the interview transcriptions based on the emerging themes, developing and revising codes as we found additional themes of parent perception and expectation. Coding was periodically cross-checked by two coders to ensure accuracy.}
    
    \subsection{Key Insights}
    \label{sec:formative_key_insights}
    
    \paragraph{\tlhighlight{\textbf{KI1:}} Parents value the educational affordance of new technologies} \yxhighlight{Nine out of the ten parents} in our sample recognized that digital technologies and AI can support their children's language and literacy learning by improving letter recognition, phoneme awareness, vocabulary, spelling, and story comprehension. Though popular press often implies that busy parents use digital devices simply as a ``babysitting'' tool, our interview indicated that parents intentionally use technologies as enriching educational opportunities for their children, especially when it concerns domains that they do not think they are capable of teaching. For example, one parent in our study said ``\textit{ (translated from Spanish) I don't speak English at all. So letting my little one watch television or use apps are important for her to learn English before school.}'' Another non-English speaking parent showed us a Luka device in their home - an AI-powered robot that can read print books aloud --commenting that they used this device as their young child's ``English learning time.'' A parent who does not speak Spanish herself mentioned that her child sometimes picked up Spanish words from talking with Alexa. The parent said  ``\textit{It's just like she can have this fun Spanish lesson with Alexa. I think, wow, that is cool.}'' Moreover, affordable digital content is a valuable learning resource for families with less access to expensive educational opportunities, such as private tutoring and enrichment camps. One parent mentioned that their child frequently played free spelling games on PBS KIDS, because \textit{``it is just available for us''}.\looseness=-1
    
    \paragraph{\tlhighlight{\textbf{KI2:}} Parents prefer interactive storytelling systems} \yxhighlight{Eight parents} mentioned that they prefer technologies that provide interactive opportunities for children, which the parents believe lead to more engaging and active learning. For example, one parent viewed positively the choose-your-own-adventure story apps on Alexa that allowed her child to control how the story proceeds by providing speech command, pointing out that \textit{``Compared to just listening to a story\ldots I think the interactive one [can] get her brain thinking, get her brain building, and working, and exercising''}.

    \paragraph{\tlhighlight{\textbf{KI3:}} Parents view technologies as a way to promote parent-child interaction} While there is a fear that children's use of technologies supplants their interaction with other family members, \yxhighlight{Six parents} in our study suggested that they think technologies can have a positive impact on family interactions, even though many of the technologies are not intentionally designed to encourage parent-child interaction. \yxhighlight{Four} parents mentioned the enjoyable moments they have had watching television or playing video games with their child, which made them ``\textit{feel closer [with their child]}'' or \textit{``like doing a family thing''}. Another parent, in particular, mentioned that some interactive features of digital books (e.g., hotspots) often triggered their child to ask her questions or make comments to her, which often turned into an interesting longer conversation. Nevertheless, parents do appreciate apps or systems that are intended for co-use, particularly those designed for educational purposes. One parent mentioned the challenges she had trying to interact with her child when using a story reading app. Although this parent was well aware of the benefits of asking children questions during reading, she told us that \textit{``I'm  trying to think of what questions I could ask my kid but sometimes I can't even think of anything off the top of my head''}.\looseness=-1 
    
    \paragraph{\tlhighlight{\textbf{KI4:}} Parents are highly involved in selecting content for their children and have a desire for customized content} We found that parents tend to carefully select digital content they think is beneficial for their children, \yxhighlight{as this theme was brought up by over half of the parents we interviewed}. They either rely on their own subjective judgment (e.g., \textit{``I just think it's good for my kid.''}) or seek out guidelines issued by researchers or institutions. For example, a parent mentioned that \textit{``I just saw this on Common Sense Media, which is where I usually go to have an initial check on age level appropriateness''}. Nevertheless, all parents expressed confidence that they know what is appropriate for their children because they know \textit{``what [their child] likes, what [their child] knows, what [their child] doesn't like.''} Therefore, it is not surprising that some parents indicated that they sometimes wish they could modify the content to better fit their child's interests or needs. For example, one parent mentioned \textit{``yeah those are good apps, but I might want to change the language a bit. I don't think my son understands this word''}. 

    \begin{figure*}[t]
        \includegraphics[width=0.8\textwidth]{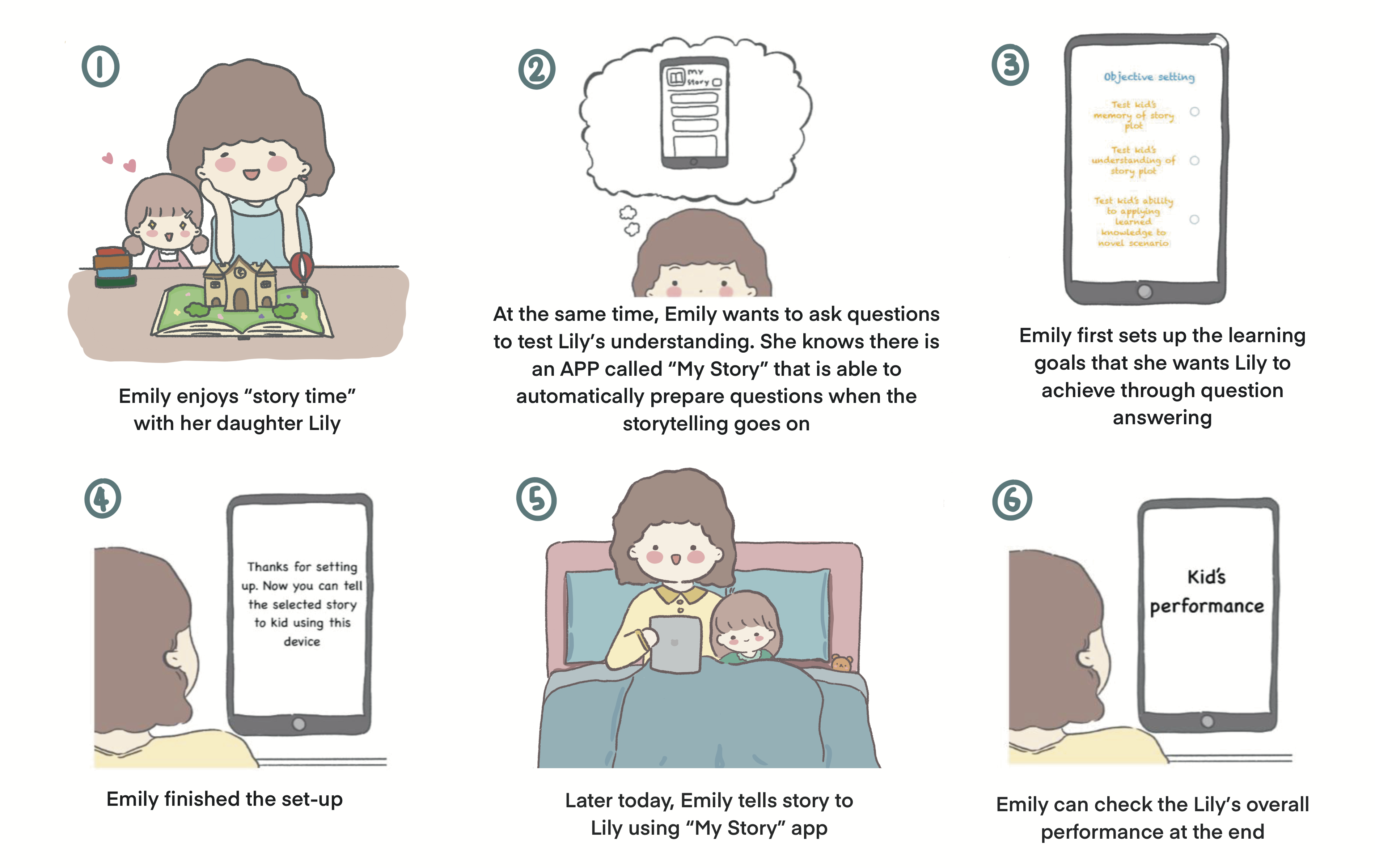}
        \caption{An example part of a storyboard used in our participatory design sessions}
        \label{fig:storyboard_example}
    \end{figure*}

\section{Participatory design}
    \subsection{Process}
        \label{sec:design_process}
        With the design opportunities identified in the formative study, we launched participatory design (PD) sessions~\cite{muller_participatory:_1993} with four parents (PA1--4) to further uncover concrete design goals and design strategies for \sys{}. 
        
        \subsubsection{Participants}
        \zzhighlight{The participants were parents recruited through the mailing lists of our maintained participant pool. Three of them spoke English as a second language but were fluent in English. Two of them were mothers of 4-year-old children and another two were mothers of children older than 5. All participants told stories frequently to their kids: Two parents told stories to their kids once a day, while another two did storytelling 4--6 times per week. All of them were primary caregivers to their kids.} 
        
        
        \subsubsection{Procedure}
        The sessions were conducted remotely via Zoom and each lasted around an hour. In each session, we presented the parent with four sets of low-fidelity storyboards, each illustrating a different scenario of using \sys{} to tell stories to their kids. The parents were asked to discuss their feelings about the scenarios in the storyboards, identify possible design opportunities, challenges, and user concerns, and ideate design strategies and new features to improve the system. Each parent was compensated with a \$25 gift card for their time.  
        
        \tlhighlight{The main goal of the PD approach is to include parents' voices and ideas in the design process, utilizing their unique experiences in helping us explore the problem space.} Prior to each PD session, we conducted a semi-structured interview with the parent to learn more about their current practice and strategies in storytelling, whether they engage in any questioning-and-answering with their children, and any challenges they encountered in storytelling.

        In each PD session, we presented the participant with four variations of storyboards in random order, an example storyboard is shown in Figure~\ref{fig:storyboard_example}. \tlhighlight{Those storyboards serve as starting points in the PD process so that participants can brainstorm new interfaces, interaction strategies, use scenarios, and system capabilities based on the variations of user needs, interaction modalities, and contexts depicted in four storyboards~\cite{muller_participatory:_1993, grudin2002ppd}. The four variations of storyboards were designed based on the key insights (Section~\ref{sec:formative_key_insights}) from the formative study as we will articulate below.}
        
        \subsubsection{Storyboards}
        \tlhighlight{
            The four variations of storyboards differed from each other in two aspects: (1) whether the parent is present at the storytelling; and (2) whether the system runs on a tablet or a smart speaker. These two aspects reflected the options in key design decisions we identified from the formative study. The storyboards also showcased the envisioned user needs and corresponding system features on interactive conversational agents, the configuration of question types, the dashboard that supports child performance tracking, and the generation of follow-up questions.}

        \tlhighlight{    
            The design of storyboards was based on Personas~\cite{grudin2002ppd} constructed using the key insights from the formative study. In particular, the parent profile in the storyboards was characterized as someone who appreciates the educational benefits in technologies for children (KI1), is willing to adopt new AI-enabled interactive systems in storytelling (KI2), values human-human interaction between parents and children (KI3), and wishes to stay involved in curating and filtering digital contents that their children consume (KI4). The parent was also sometimes ``busy and unavailable'' and therefore absent from the synchronous storytelling session in the ``parent absent'' variations of storyboards. The use of persona as a PD tool helps users understand the design challenges and concretize their ideas~\cite{grudin2002ppd}, which has also been previously used in similar design domains~\cite{RUBEGNI2022102727, Alexandrakis2019Insights, Sylla2013Design}.  
        }

        The storyboards set scenarios for users to reflect on their needs, constraints, and practices~\cite{bodker1999scenarios}. In the storyboards, we purposely deemphasized the details in the interface designs of the system by avoiding directly showing screen contents (if showing screens was necessary, our storyboards used low-fidelity sketches). Instead, the storyboards focused on illustrating the parent motivations and goals in the scenarios, the constraints in time, attention, and cognitive capabilities, and the interaction dynamic among the parent, child, and agent in different scenarios depending on whether the parent is present and which type of device is used. The main goal was to elicit the feelings and emotions of participants towards different design decisions and to build empathy with them. After validating user needs illustrated in the storyboards, we asked participants to think about how their personal experience with parent-child storytelling (or the lack of it) can connect to the storyboard scenarios. Through this process, each parent identified things they liked, things they disliked, and their concerns about the different paradigms of AI involvement in the parent-child interactive storytelling process. Lastly, we asked them to think of and propose new ideas on design features, interfaces, or interaction techniques for (1) addressing the issues they identified in the storyboards; and (2) bridging the gaps between the scenarios presented in the storyboards and their own personal scenarios.

    \subsection{Findings}
    
    
    \subsubsection{\tlhighlight{Optimizing for engagement as a key goal in storytelling}}
    \label{sec:engagement_goal}
    \tlhighlight{Achieving high child engagement in storytelling is a key goal for parents in both kinds of scenarios: when the parent is present and when the parent is absent.} When a parent participates in the storytelling, maintaining child attention is challenging as children quickly get bored and subsequently distracted when there is a lack of change in the type of activities or interaction patterns. This issue becomes more problematic when the parent is not present. When encountering time and attention conflicts, parents often seek to use digital content such as videos of stories and songs on smartphones to keep children occupied when parents are in meetings (especially common during the COVID-19 pandemic when many parents work from home and have remote meetings) or doing housework. However, children quickly get bored and try to seek attention from their parents when they are unavailable, resulting in frustration for both parties (PA1, PA2).

    When we asked about practical strategies that participants currently used to enhance children's engagement in storytelling (without the involvement of an intelligent agent), joint reading came up as a common strategy that parents usually adopted: ``\textit{Usually, we'll take turns since she doesn't like to read all the pages by herself. So, so I want to, like motivate her. And usually, I'll read one sentence and she will read the next sentence and we will take turns to read the whole book.}'' (PA2), ``\textit{So we do what's called joint reading. So I'll read, and then I'll have him read. And then I'll read and we go back and forth.}'' (PA3). In addition to enhancing engagement, some participants also valued the effect of joint reading on providing emotional support to their kids: ``\textit{in addition to engagement, I think, (kids will get) the emotional support, while you're reading with your child, and the interaction with the child, especially before (going to) bed, \ldots your child will feel very loved and warm}'' (PA2).
    
    
    \tlhighlight{The PD sessions identified providing multiple variations of interaction patterns as a key strategy for improving engagement.} An ideal digital system for interactive storytelling should support (1) flexible switching between who is reading and who is facilitating the questions among the agent, the parent, and the child; (2) diverse question-answering patterns with various question types, follow-up questions on correct answers, and guidance for incorrect answers. 
    
    \tlhighlight{An ideal digital storytelling system should also balance the parent's desire for involvement (as reported in Section~\ref{sec:formative_study}) and their practical need for minimizing parental intervention needed when busy by providing distinct modes.} In a parent-agent co-reading mode, the agent should play a \textit{supportive} role that helps the parent identify opportunities for questions, recommends appropriate questions, and offers options for occasional child-bot interaction to make the process more ``fun'' and engaging for the child. While in a parent-absence mode, the agent needs to play a \textit{proactive} role in engaging the child with the goal of minimizing their need for parental attention. Nevertheless, despite the lack of \textit{synchronous} parent involvement at the time of storytelling in this mode, many parents still desire \textit{asynchronous} parent involvement through configuring the agent's interaction plan beforehand and tracking the child's progress afterward.

    \subsubsection{\tlhighlight{Challenges and opportunities of question-answering in digital storytelling}}
    \label{sec:challenges_qa}

    In current joint-reading practice, all four participants used questioning-and-answering to (1) improve engagement with children; and (2) assess and develop their comprehension in the storytelling. The participants often asked simple questions whose answers were apparent in the story, such as simple math questions (PA1), questions about color or shade (PA1, PA4), questions about pictures (PA2), or questions about major actions in the story (PA3). However, participants also  recognized the importance of asking questions of different varieties and reported that children might have different demands for questions as they grow up: ``\textit{I think it (question type) depends on age, like, when kids are smaller, ... they want someone to ask very specific questions. And maybe for other kids (who are older), they want more challenging questions, or they don't want someone to interrupt them during the reading, instead, they prefer to answer the question afterward or before reading}'' (PA2) 
    
    Coming up with appropriate questions of different types, especially in real-time while reading the story and interacting with the child can be challenging. As reported by PA4, sometimes she ends up finishing reading a story without asking questions due to the limitation in her cognitive load despite knowing about the benefits of question-answering. 
    
    \subsubsection{\tlhighlight{Varied assessment goals of parents in question-answering}}
    \label{sec:varied_goals}
    
    \tlhighlight{
    Among four participants, PA3 was the only one who explicitly reported using question-answering as an assessment tool (``\textit{to make sure they are understanding the facts''}). She often liked to her child follow-up questions on rationale (e.g., why is that?) and emotion (e.g., how do you think Susie will feel?) after questions about facts in the story plots (e.g., what happened\ldots?) in order to assess the development of different skills of her child. Other parents also reported asking follow-up questions but as a means for maintaining child engagement instead of assessment.}

    \tlhighlight{
    Parents liked the idea of using AI assistance for identifying opportunities of asking questions and generating possible questions to use during parent-child joint reading. In the PD process, both PA2 and PA3 recommended design strategies for grouping similar questions together by themes, relevant entities, or question types to make it easier for parents to plan for them. }

    \tlhighlight{
    However, parents had diverging opinions on the assessment-focused features of the system. While PA3 was quite excited about the idea and suggested how assessment goals can be grouped into testing specific capabilities of the child, PA4 was concerned about whether the assessment goals embedded in the generated questions would affect the child's interest in the storytelling agent. She expected that the child would dislike the agent once they discovered that the agent was trying to assess them.}

    \subsubsection{\tlhighlight{Balancing between desires and constraints in the granularity and form of parental involvement}}
    \label{sec:conflict_parental_involvement}
    Results from our PD sessions confirmed findings from the formative study that parents wish to have active involvement in the selection and configuration of contents in AI-facilitated storytelling, but the desired degree of involvement varies. PA1, PA2, and PA4 were OK with configuring question types, but were not too keen on the idea of editing and tweaking each individual question recommended by the system. However, PA3 reported her desire for controlling questions at a fine level of granularity despite that she did not expect most other parents to like it. PA3 said \textit{``I think this [configuring individual questions] will overwhelm parents\ldots[parents] want to rely on the app to figure out what those questions are. And you know, that you don't want to have to think about it\ldots I would definitely love this, I think it could be an option.''} To address this divergence in user needs, our system should support flexible levels of parent control in story and question contents.
    
    \tlhighlight{When it comes to the parent involvement in the delivery of contents, a conflict arises between (1) the parent's desire to be present, play an active role and foster the parent-child relationship; and (2) the constraint that sometimes they are not available is present for all four participants, which is consistent with the findings in the formative study (Section~\ref{sec:formative_study})}. Parents liked the fact that two variants of the system described in the storyboard can facilitate the storytelling process and the interaction with the child on its own without parent intervention, while still allowing asynchronous involvement from parents by controlling the story and question content beforehand and tracking the child's progress through a dashboard afterward.


    \subsection{Design Strategies}
    
    The PD process helped us identify the following six key design strategies, \tlhighlight{which will guide the design of the system described in Section~\ref{sec:storybuddy}}:
    \begin{itemize}
        \item \textbf{DS1}: Maintain child attention and optimize for child engagement through the alternating use of many variations of interaction means and approaches among the parent, the child, and the storytelling agent \tlhighlight{(Section~\ref{sec:engagement_goal})}.
        \item \textbf{DS2}: Assist parents with facilitating question-answering in joint-reading through recommendations of questions to ask and opportunities to ask questions \tlhighlight{(Section~\ref{sec:challenges_qa})}.
        \item \textbf{DS3}: Support different parental goals on whether to focus on the assessment objective in interactive storytelling \tlhighlight{(Section~\ref{sec:varied_goals})}.
        \item \textbf{DS4}: Generate appropriate follow-up questions with relevant semantic entities or story themes to improve child engagement and the effectiveness of assessments \tlhighlight{(Section~\ref{sec:varied_goals})}.
        \item \textbf{DS5}: Accommodate varied parent preferences on the granularity to configure interaction contents through providing multiple flexible options \tlhighlight{(Section~\ref{sec:conflict_parental_involvement})}. 
        \item \textbf{DS6}: Address the conflict between (1) parents' desire to be present and involved in the live storytelling process in order to strengthen parent-child relationships and (2) the constraint that sometimes they are not available by supporting both (1) synchronous parent-child joint-reading with AI assistance and (2) AI-facilitated storytelling where question-answering contents were asynchronously configured by parents \tlhighlight{(Section~\ref{sec:conflict_parental_involvement})}. 
    \end{itemize}

\section{StoryBuddy}
\label{sec:storybuddy}
    \begin{figure*}[t]
        \includegraphics[width=0.7\textwidth]{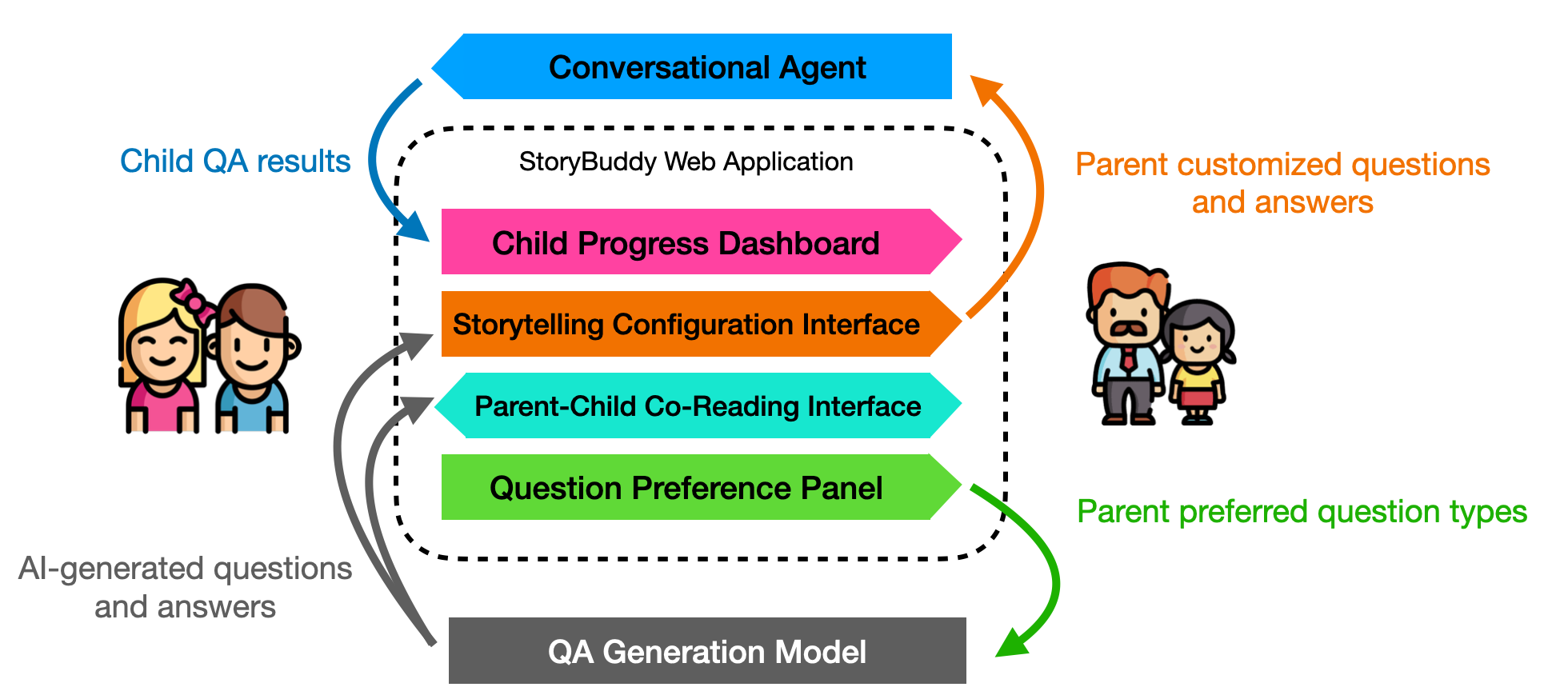}
        \caption{The system architecture of StoryBuddy. Lines with arrows represent data exchange between modules. The directions of the pointers on modules indicate whether the module interacts with the parents (right) or the children (left).}
        \label{fig:architecture}
    \end{figure*}
    
    \subsection{System Overview}
    Following these six design strategies, we designed and implemented StoryBuddy, an AI-enabled interactive tool for configuring, augmenting, and automating interactive storytelling with children. StoryBuddy presents several features that allow flexible parent involvement in both the configuration and delivery of interactive story contents, while supporting diverse parent needs in children's skill development, progress assessment, and engagement.
    
    As shown in Figure~\ref{fig:architecture}, StoryBuddy consists of: (1) a storytelling configuration interface for parents to configure the question answering contents; (2) a parent-child co-reading interface for assisting the parent with the joint-reading process; (3) a conversational agent that can coordinate question-answering and automate storytelling when the parent is absent; (4) a dashboard that tracks and displays the child's progress; and (5) a back-end machine learning model that can generate possible questions and answers for \textit{any} story.
    
    \subsection{Modes for Parent Presence and Absence}
    \tlhighlight{Informed by design strategies DS3, DS5, and DS6 from the formative study and PD sessions, we decided to create two distinct modes in StoryBuddy to reconcile the parent's desire to be present, play an active role, and strengthen the parent-child relationship with the constraint that they are sometimes not available for live storytelling.} 
    
    \begin{figure*}[t]
        \includegraphics[width=0.7\textwidth]{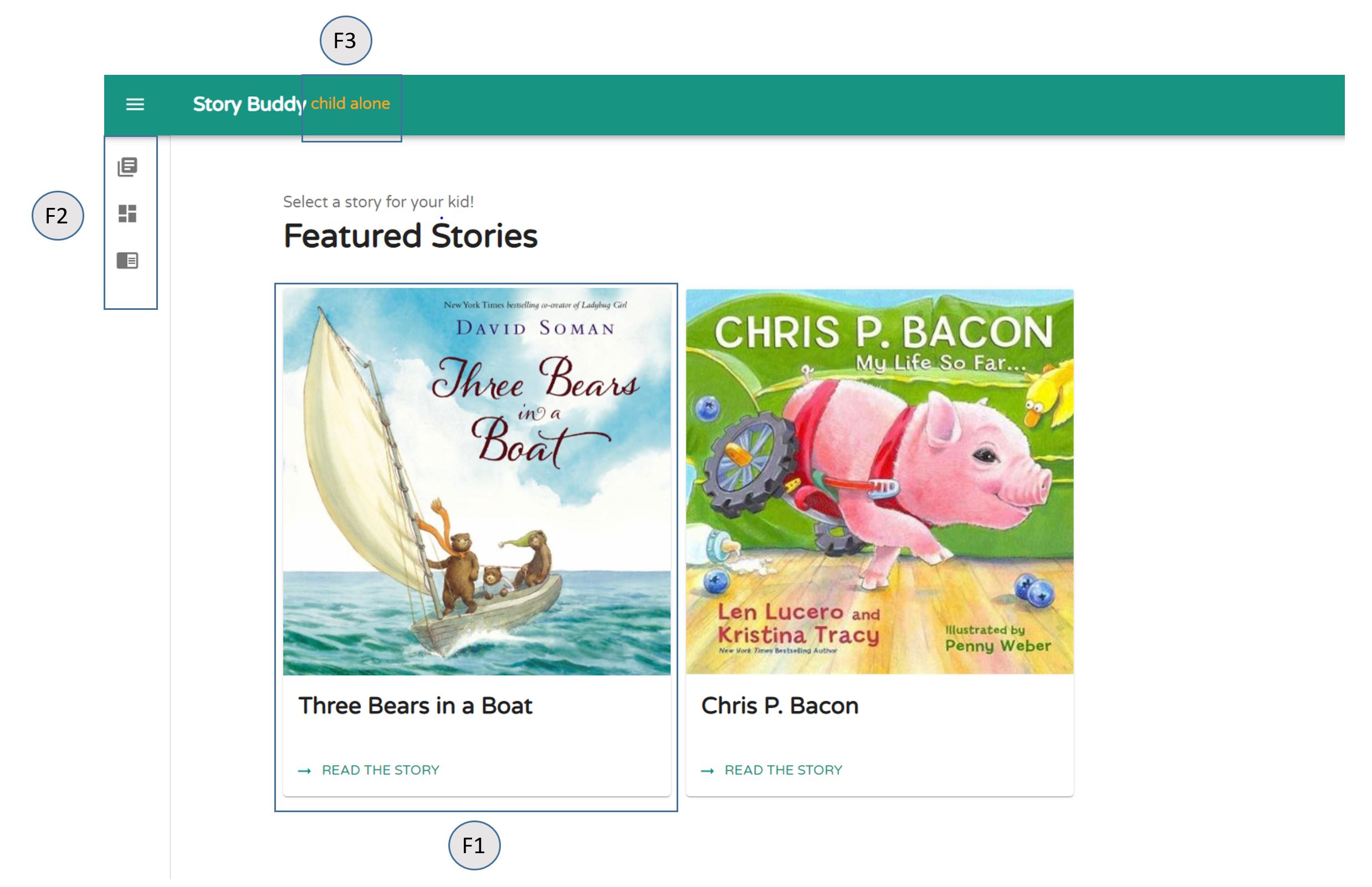}
        \caption{The story library panel in StoryBuddy}
        \label{fig:storySelection}
    \end{figure*}
    
    \subsubsection{Parent presence}
    \label{sec:parent_presence}
    \tlhighlight{As identified in the PD sessions, when the parent is present, the main goal of StoryBuddy is to assist the parent by helping them identify opportunities to ask questions and recommend questions to use in order to help reduce their cognitive load~\tlhighlight{(DS2)}, while at the same time augmenting the parent by providing new variations of interactions for the child so they can interact with a conversational bot \textit{in addition} to their parent to enhance the child's engagement~\tlhighlight{(DS1)}}. It is important that StoryBuddy does not \textit{displace} or \textit{lessen} the parent's role, preserving the parent-child relationship-building aspect of storytelling that both the parent and child treasure~\tlhighlight{(DS6)}.
    
    To use StoryBuddy in the parent-child joint reading mode, the parent can simply choose a story from the story library panel (Figure~\ref{fig:storySelection}). and enter the story reading interface (Figure~\ref{fig:parentReading}). The story content panel on the left displays story text (F4) and the corresponding illustration (F5) of the current page. Parents can navigate through the pages by clicking left and right button. 
    
    On the right side, there is the question panel (F7 and F8 in Figure~\ref{fig:parentReading}) that displays the AI-generated recommended questions, when the parent selects a question, the corresponding part of the story is highlighted, indicating the connection between questions and story contents. The parent can read the story to the child using the story content panel first, and decide if they want to ask any of the recommended questions from the question panel. They may click on the question so it expands to show the correct answer, and click on either the check or the cross button to record the correctness of the child's answer to be aggregated in the dashboard (see Section~\ref{sec:dashboard}). Clicking on the check or the cross button triggers the generation of a follow-up question. The follow-up question will be about a relevant entity or a different aspect of the same entity in the original question (see Section~\ref{sec:question_generation}).
    
    \tlhighlight{An objective of our design of the parent-child joint reading mode is to give the parent a maximum level of control and agency~(DS5)}. If they like, they could handle almost all aspects of the storytelling themselves without taking advantage of any ``smart'' features. However, they could also feel free to use the recommended questions and the generated follow-up questions as they see fit. \tlhighlight{They might also delegate question-asking, answer-checking, progress-tracking, and even the reading of the story itself to the agent as they wish~(DS2). These features correspond to the PD finding that parents wish to have flexible degrees and granularity of control of their involvement in the digital storytelling experience.} 
    
    \paragraph{Dynamic interaction paradigms for engagement}
    Maintaining child engagement is an important goal in parent-child joint reading mode. \tlhighlight{As identified in the PD, a potentially effective strategy is for StoryBuddy to support dynamic interaction paradigms so that the parent and the child can switch between different ways of interaction~(DS1)}. This way, the child does not get easily bored. In parent-child joint reading mode, the parent can change the way of interaction on three aspects: (1) While the default setting is for the parent to read the story, the parent may easily have the agent read the story by clicking on the play icon, as shown in Figure~\ref{fig:parentReading}. (2) After the parent asks a question, the child can either answer the question by speaking to the parent who will manually check the correctness of the answer using the question panel, or speaking to the agent by clicking on the microphone icon, as shown in Figure~\ref{fig:parentReading}. When the child speaks to the agent, the agent can judge the correctness of the answer and further engage with the child (more details discussed in Section~\ref{sec:parent_absence} about the automated bot-reading mode). (3) The parent may also quickly invoke the conversational agent to handle follow-up questions on their behalves in the parent-child joint reading mode by switching to the chatbot panel (F6 in Figure~\ref{fig:parentReading}).

     \begin{figure*}[t]
        \includegraphics[width=0.7\textwidth]{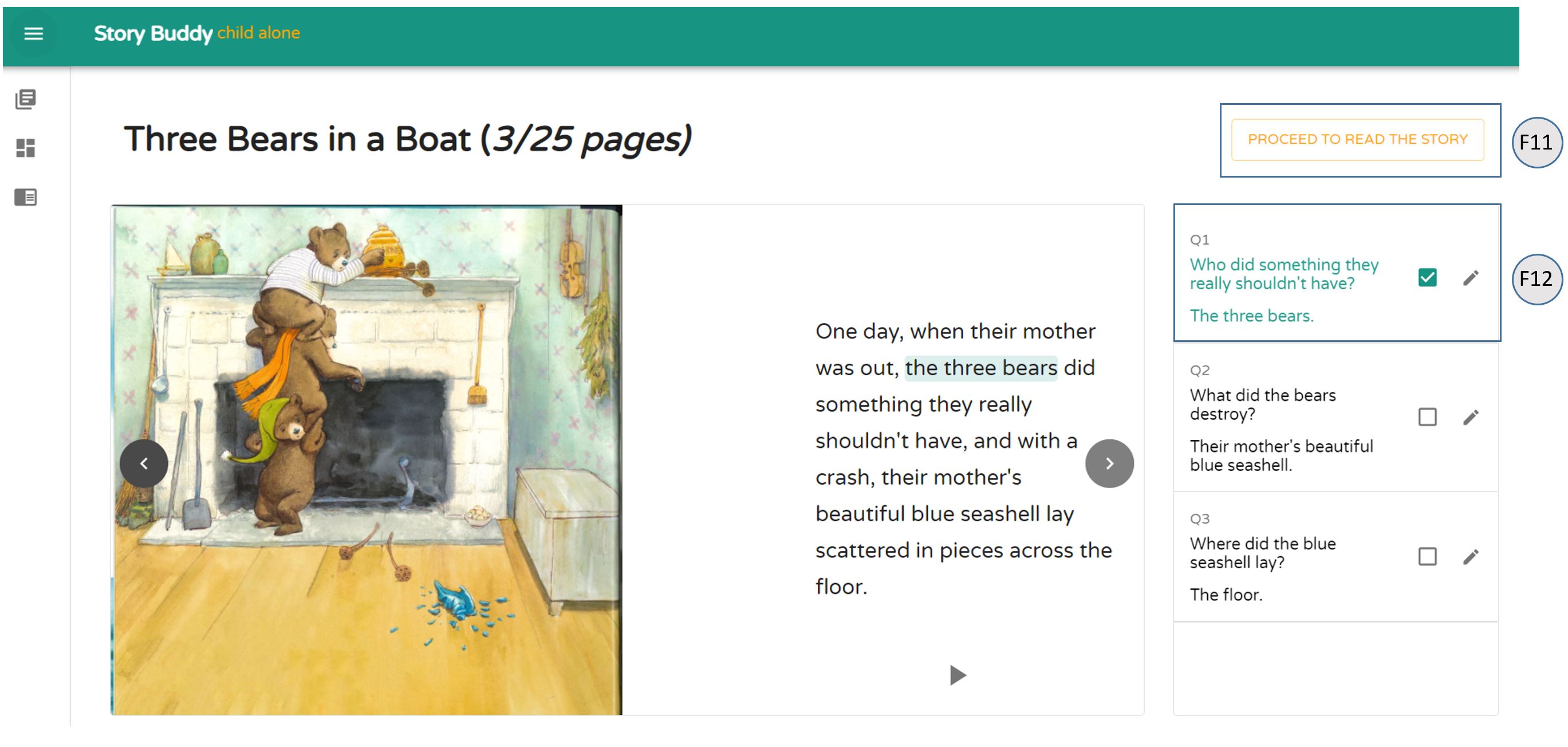}
        \caption{StoryBuddy's configuration page for questions to use in the automated bot-reading mode when the parent is absent}
        \label{fig:parentPreconfiguration}
    \end{figure*}
    
    \begin{figure*}[t]
        \includegraphics[width=0.7\textwidth]{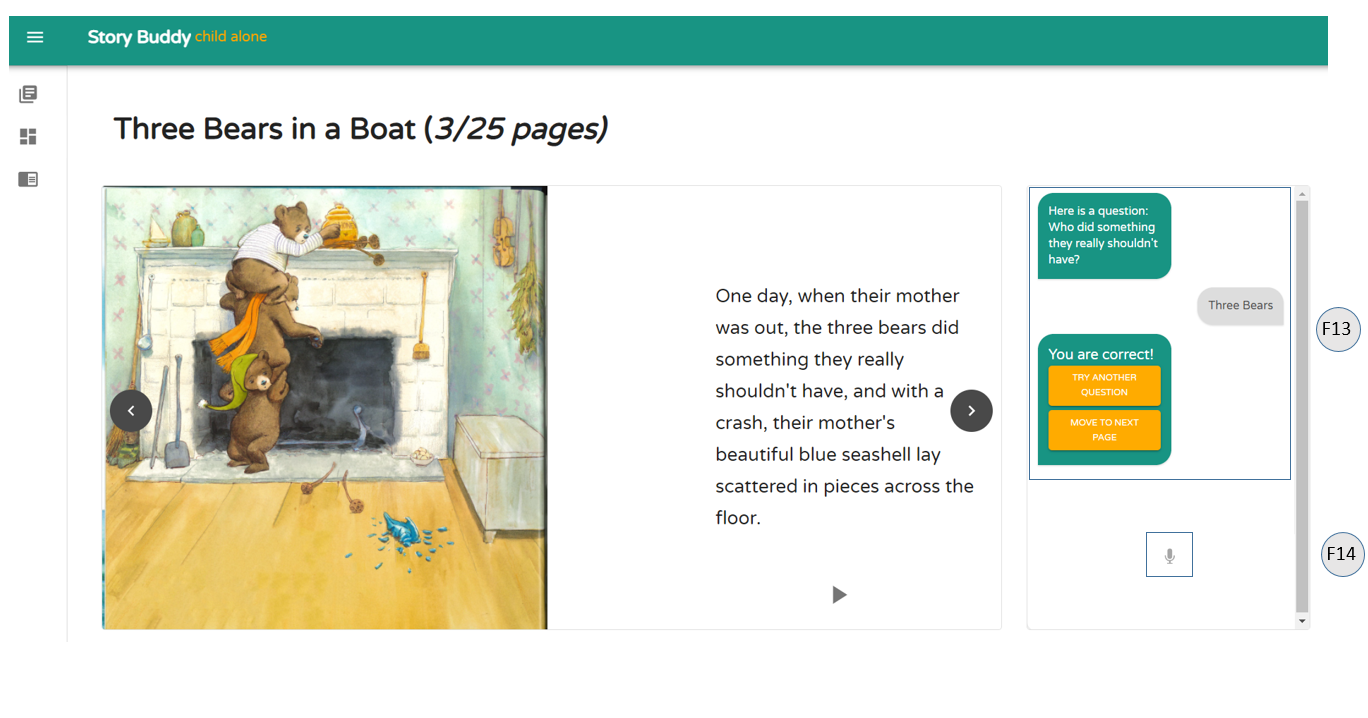}
        \caption{StoryBuddy's story reading interface in the automated bot-reading mode}
        \label{fig:childAlone}
    \end{figure*}

    \subsubsection{Parent absence}
    \label{sec:parent_absence}
    When the parent is absent, StoryBuddy operates in an automated bot-reading mode. \tlhighlight{The main goal of StoryBuddy in this mode is to engage children in interactive storytelling without requiring parents sitting next to the child during the storytelling~(DS1, DS6)}. However, while parents are absent for the \textit{synchronous} content delivery stage, often due to them being busy with other things, parents can still stay involved \textit{asynchronously} through the configuration of the StoryBuddy agent and the tracking and assessment of child progress~\tlhighlight{(DS4, DS5, dashboard details discussed in Section~\ref{sec:dashboard})}.
    
    In this automated bot-reading mode, parents can configure how the StoryBuddy agent interacts with their child in advance by customizing the questions inserted in the stories. Similar to the parent-child joint-reading mode, the parent selects a story from the story library panel (Figure~\ref{fig:storySelection}). They then enter the configuration page (Figure~\ref{fig:parentPreconfiguration}). The configuration page looks similar to the reading page in the previous mode, except for in the question panel, the parent can see all AI-generated questions and their corresponding follow-up questions in a list. The parent can choose which questions should be asked by the agent using the checkbox. In addition, the parent can edit the AI-generated questions and answers by clicking on the pen icon. This configuration step is optional--parents may directly click on the ``\textit{Proceed to read the story}'' option to skip the rest of the configuration process. By default, StoryBuddy selects the top-ranked AI-generated question and its follow-up question for each page of the storybook.
    
    \paragraph{Child-Agent Interaction in StoryBuddy}
    At storytelling time, children can see the book page including text and illustrations as shown in Figure~\ref{fig:childAlone}. On the first page, the agent first greets the child. The agent will then read the story text on each page, say ``\textit{OK, here is a question}'' and then ask the question as configured by the parent. The child can answer the question to the bot by clicking on the microphone icon (F14), the transcript of their speech will automatically appear in the dialog (F13). After receiving the child's answer, the agent will judge the correctness of the answer (technical detail in Section~\ref{sec:implementation}). If the answer is correct, the agent will say ``\textit{You are correct! Good job!}'', the child can choose either ``\textit{move to next page}'' or ``\textit{try another question}'' (if another question is available); if the answer is wrong, the child will see an additional ``\textit{try again}'' option for them to retry the same question.

    \begin{figure*}[t]
        \includegraphics[width=0.7\textwidth]{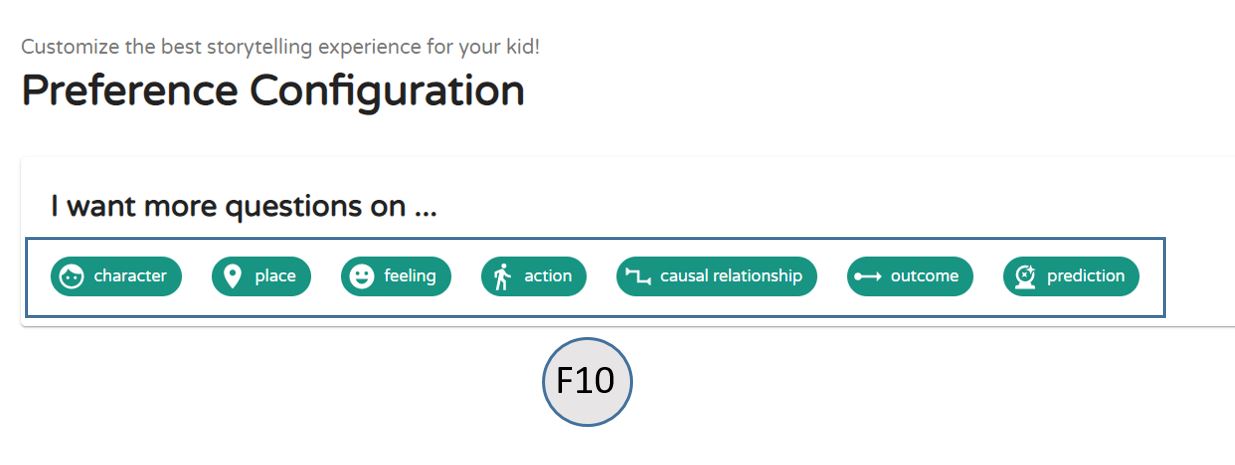}
        \caption{StoryBuddy's interface for question preferences configuration}
        \label{fig:questionConfig}
    \end{figure*}

     \subsection{Preference Configuration of Question Types}
     \label{sec:question_preference}
     StoryBuddy provides a preference configuration panel (Figure~\ref{fig:questionConfig}), where the parent can choose the preferred types of generated questions for the back-end model. \tlhighlight{From the PD insights, we learned that some parents wish to have controls of the generated questions at a finer granularity~(DS5).} The preference configuration panel allows them to customize the generation of questions to better align with the learning and development goals they have for their children. The use of this panel is optional. 
     
     StoryBuddy allows the parent to indicate their preferences for questions focusing on seven different narrative elements, including questions about story \textit{characters, setting, feeling, actions, causal relationships, outcomes, and predictions of future events}~\cite{paris2003assessing}. Character questions either start with ``who'' and ask the child to identify a character in the story, or ask the child to use information in the story to describe the character (e.g., ``How did the man's daughter look?'' ). Setting questions typically start with ``where'' or ``when'' and focus on a place or time that story events take place. Feeling questions ask the child to describe the emotion that characters  experience (e.g., ``How did the princess feel in her new home?'') Action questions are typically phrased as ``what does somebody do'' or ``how does somebody do something'', for which the child needs to provide an answer that contains certain actions (e.g., ``What did the cook do after she opened the hamper?'' or ``How did the prince break the curse on the princess?''). Causal relationship questions start with ``Why'' or ``What makes\ldots'' that ask the child to identify the causes of a focal event in the story. Outcome questions ask the child to describe the outcomes or consequences of a focal event. Lastly, prediction questions ask the child to think about what might happen next (e.g., ``How will the other animals treat the duckling?'')

   \begin{figure*}[t]
        \includegraphics[width=0.7\textwidth]{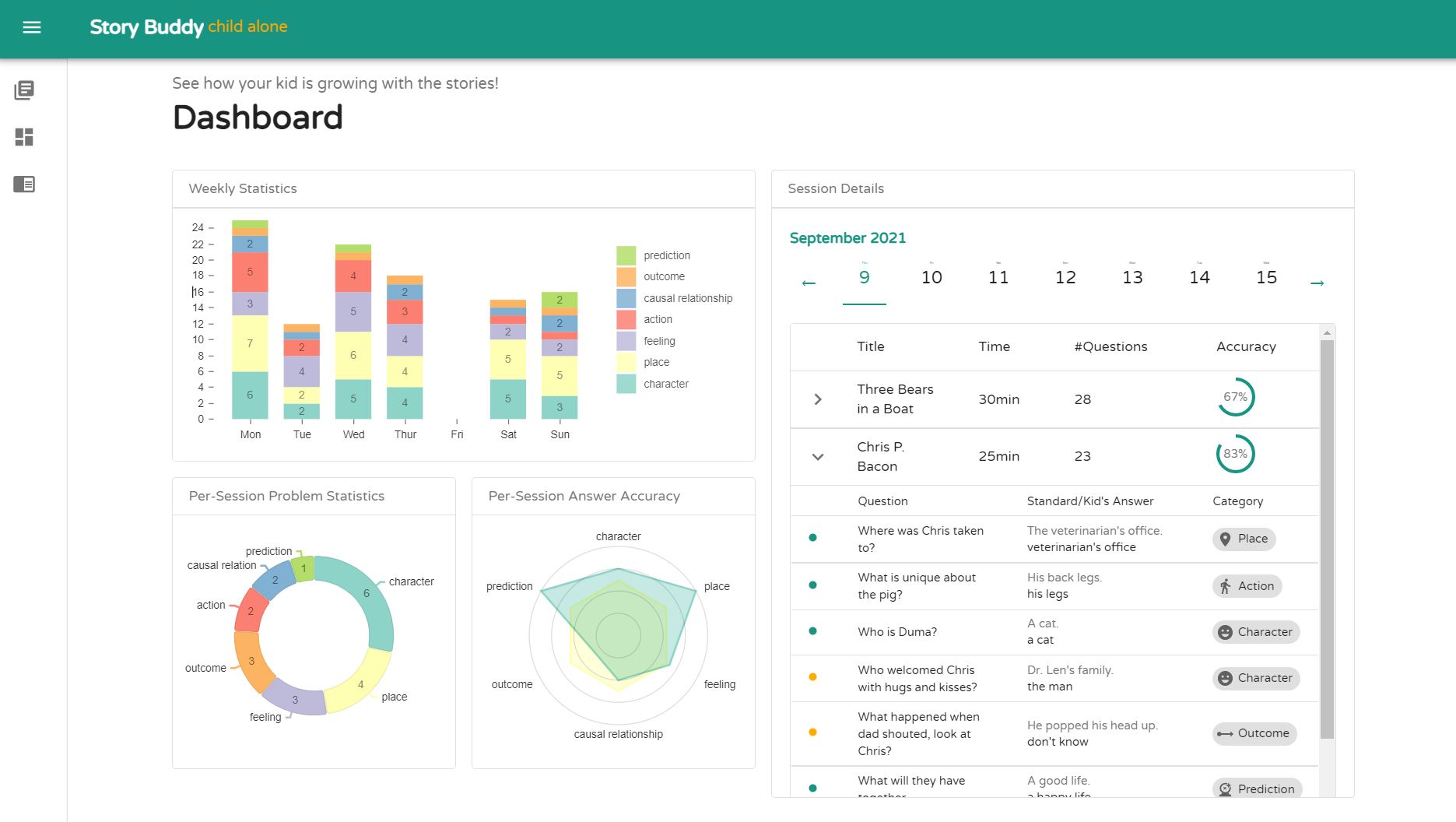}
        \caption{StoryBuddy's dashboard for child progress tracking and performance assessment}
        \label{fig:dashboard}
    \end{figure*}
    
     \subsection{Dashboard for Child Progress Tracking and Performance Assessment}
     \label{sec:dashboard}
     \tlhighlight{Another need of some (but not all) parents discovered from the PD insights is to track their child's progress and assess their child's performance~(DS4)}. To address this, StoryBuddy provides an interactive dashboard (Figure \ref{fig:dashboard}). The dashboard can show either the child's performance in a particular individual storytelling session or the child's aggregated performance over a week. As shown in Figure \ref{fig:dashboard}, when the parent clicks on a previous storytelling session, the dashboard will show the parent information regarding each question that the child tried to answer in this session (F15), including the child's attempts and the right answer to each question. StoryBuddy also shows the child's overall accuracy, their accuracy on type of questions, and the proportion of each question type in this session (F17).

     In addition, the parent can check the child's weekly progress. As shown in Figure~\ref{fig:dashboard}, the dashboard allows them to review the child's weekly progress and the overall accuracy in question-answering (F16). They can also tailor the dashboard to display the statistics of a particular question type. Besides, the dashboard also informs the parent of the proportion of each question type from all sessions in the week (F16).

\subsection{Implementation}
\label{sec:implementation}
        \subsubsection{Web Application}
        The front-end interactive web application of StoryBuddy is implemented in React and hosted using Python's built-in HTTP server. The web-based nature of StoryBuddy allows it to run from the web browsers on a variety of devices including desktops, laptops, tablets, and smartphones. The use of React allows it to be ``responsive'' so that its graphical user interfaces can adjust to fit different screen dimensions and ratios. For all functionalities to work properly, StoryBuddy requires the device to have a microphone and a speaker.
        
        StoryBuddy uses Google's Cloud Text-to-Speech API\footnote{https://cloud.google.com/text-to-speech} for speech synthesis in story-reading, which yields natural-sounding results. Storybooks in StoryBuddy are stored in a simple JSON format, which allows users and community members to easily add new storybooks to its story library.
        
        
        
        \subsubsection{Conversational Agent}
        The conversational agent used in facilitating agent-child reading is implemented with the \textit{react-simple-chatbot} framework\footnote{https://github.com/LucasBassetti/react-simple-chatbot} at the front end for displaying the chat history and facilitating input/output. At the back-end, it uses the Google Dialogflow \footnote{https://cloud.google.com/dialogflow} framework for intent detection and classifying the child's answers to determine their correctness. We trained the answer classification model in Dialogflow with a small rule-based corpus. Given a model-generated answer (Section \ref{sec:question_generation}), we proliferated the Dialogflow training phrases of each question by applying templates upon its answer, so that the agent can correctly handle variations in the answer (e.g., ``three bears'' vs. ``3 bears'') as well as fillers in the answer such as ``\textit{It may be <answer>}'', ``\textit{I believe <answer>}'', and ``\textit{I guess <answer>}''. Note that the Dialogflow training only takes a few seconds, therefore when a new question is generated, the chatbot becomes ready to respond to the user's answer in real-time. \looseness=-1

        \subsubsection{Question Generation Model}
        \label{sec:question_generation}
        For question generation, StoryBuddy uses an automated question-answer generation (QAG) model trained on the FairytaleQA dataset~\cite{yao2021ais}. This QAG model can automatically generate high-quality QA pairs from any children's storybooks. The questions generated by the QAG model are designed to mimic the style ``as if a teacher or parent is to think of a question to improve children's language comprehension ability while reading a story to them~\cite{yao2021ais}''. This QAG model also supports generating questions and the corresponding answers from a specific question type (detail of the supported types in Section~\ref{sec:question_preference}).  
        
        On a high level, the pipeline of this QAG model consists of a rule-based answer generation module, a BART-based question generation module~\cite{lewis2019bart}, and a ranking module. The QAG model was trained on the FaiytaleQA dataset, which contains 922 QA pairs from 46 children's storybooks labeled by education experts. In a human evaluation, this QAG model achieved state-of-art performance in generating high-quality question-answer pairs from children's storybooks~\cite{yao2021ais}.
    
        In order to identify follow-up questions, StoryBuddy groups a large amount of QAG-generated questions for each section. The strategy that StoryBuddy uses is to treat the top-3 questions as anchors and then calculate the similarity between each of the remaining questions and the anchored questions. We define similarity as the number of overlapping tokens after removing the stop words. A question will become a candidate for the follow-up question of an anchored question if the similarity between them is greater than 3 and the answer of the question is not included in the anchored question text. An anchored question will not have a follow-up if it lacks an eligible candidate.

\section{User Study}

We conducted a remote user study\footnote{The study protocol was approved by the IRB at our institution.} to evaluate StoryBuddy. The study examined the following research questions: 

\begin{itemize}
    \item \textbf{RQ1}: Can parents successfully use StoryBuddy to create interactive storytelling experiences for their children?
    \item \textbf{RQ2}: How do parents and children interact with StoryBuddy in its two modes?
    \item \textbf{RQ3}: Do parents and children find StoryBuddy usable, useful, and likable?
\end{itemize}

\begin{table*}
    \def\arraystretch{1.1}
    \centering
        \begin{tabular}{ |c|c|c|c|c|c|c|  }
             \hline
             \textbf{ID} &\textbf{Parent's gender}&\textbf{Child's gender}& \textbf{Parent's age} & \textbf{Child's age} &  \textbf{Storytelling frequency} & \textbf{ESL?} \\
             \hline
             PB1 & Female & Female & 45-54 & 5 & More than 6 times a week & N \\ 
             PB2 & Female & Female & 35-44 & 7 & More than 6 times a week & N \\ 
             PB3 & Male & Male & 25-34 & 5 & More than 6 times a week & Y \\ 
             PB4 & Female & Female & 25-34 & 4 & 1-3 times a week & N \\ 
             PB5 & Female & Female & 35-44 & 7 & 4-6 times a week & Y \\ 
             PB6 & Female & Male & 25-34 & 4 & More than 6 times a week & Y \\ 
             PB7 & Female & Female & 25-34 & 7 & More than 6 times a week & Y \\ 
             PB8 & Female & Male & 25-34 & 5 & 1-3 times a week & Y \\ 
             PB9 & Male & Male & 25-34 & 5 & 1-3 times a week & Y \\ 
             PB10 & Female & Female & 35-44 & 7 & Not reported & N \\
             PB11 & Female & Female & 35-44 & 6 & More than 6 times a week & Y \\ 
             PB12 & Female & Female & 35-44 & 6 & 4-6 times a week & Y \\ \hline
        \end{tabular}
    \centering
    \caption{Demographics of user study participants.} 
    \vspace{-0.4cm}
    \label{tab:demographics}
\end{table*}

\subsection{Participants}
We recruited 12 pairs of participants (PB1--PB12) from university mailing lists and through the snowball sampling method \cite{Naderifar2017SnowballSA}. Each pair consisted of a parent and a child between the ages of 3--8. All the participants resided in the U.S. and were fluent in English. \zzhighlight{Eight parents used English as their second language.} The demographic characteristics of the participants are reported in Table~\ref{tab:demographics}.  Each pair of participants was compensated with a \$50 gift card for their time.

\subsection{Study Procedure}
Each user study session lasted around an hour and was conducted remotely over Zoom due to the impact of the COVID-19 global pandemic. Participants accessed StoryBuddy using the browser on their own computers and shared their screens with the experimenter. Participants were also encouraged to turn on their cameras if possible. All user study sessions are video recorded.  

Prior to the beginning of each session, the parent signed the consent form and filled out a demographic questionnaire. After the experimenter gave a short introduction to the study, the participants watched a 4-minute tutorial video on how to use StoryBuddy.

\begin{figure}[t]
        \includegraphics[width=\linewidth]{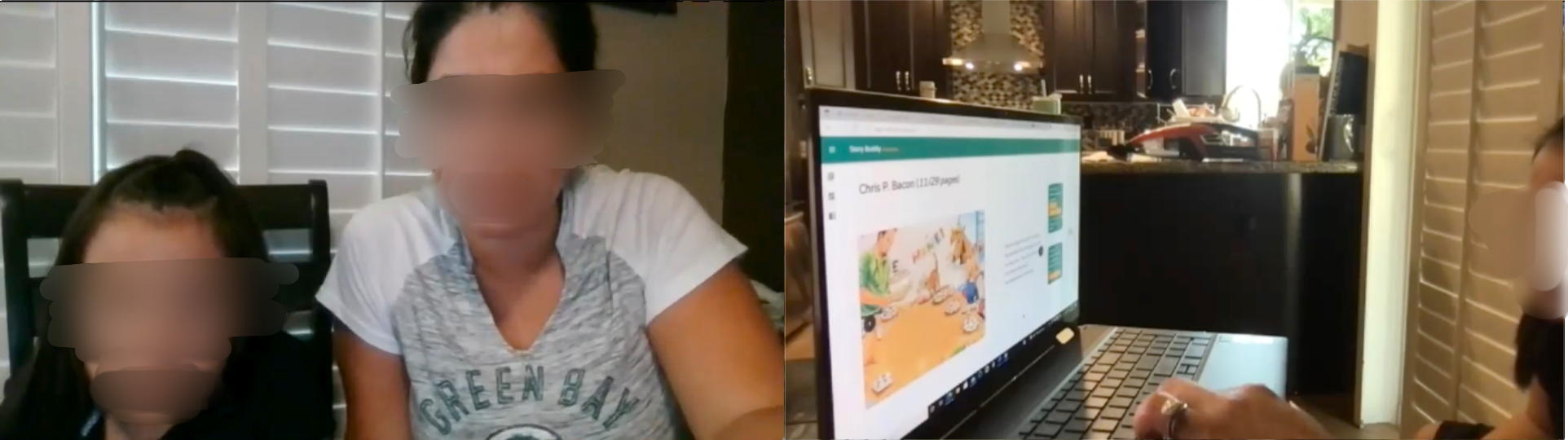}
        \caption{A screenshot from a remote user study session, showing a child and her parent interacting with StoryBuddy}
        \label{fig:user_study}
    \end{figure}

Each pair of participants then used StoryBuddy to read two stories: \textit{Three Little Bears} and \textit{Chris P. Bacon: My Life So Far} in two modes (\textit{parent-AI co-reading} and \textit{automated bot-reading}). The two stories were chosen because of their appropriate lengths for the study and appropriate difficulties for the target age group. The order of the stories, as well as the match between the modes and the stories, were random. In the parent-AI co-reading mode, the parent read the story to their child using the story reading interface and facilitated interactive question-answering with the help of StoryBuddy, as described in Section~\ref{sec:parent_presence}. In the automated bot-reading mode, the parent first customized the questions to be used by the agent using the configuration page. They then used the StoryBuddy agent to automatically read the story and interact with the child with questions, as described in Section~\ref{sec:parent_absence}. We asked the parent to avoid intervening when the child was interacting with the agent in the automated bot-reading mode. Figure \ref{fig:user_study} illustrated the scenario of how the parent and the child interacted StoryBuddy in our study: the parent sat beside the child and they used StoryBuddy to read the assigned story in one of the modes. They could chat with each other during the study.

After trying out StoryBuddy, we conducted a 10-minute semi-structured interview with each participant on their experience interacting with StoryBuddy.

    \subsection{Study Results}
    
    All 12 pairs of participants successfully completed the assigned parent-AI co-reading and automated bot-reading sessions. The parent-AI co-reading session and the automated bot-reading session lasted around 18 and 17 minutes on average respectively.

    \begin{table*}
    \def\arraystretch{1.2}
    \centering
        \begin{tabular}{ |p{5cm} | p{9cm}|  }
             \hline
             \textbf{Research questions} & \textbf{Behaviors of interest} \\
             \hline
             \multirow{3}{*}{\parbox{5cm}{How did parents read story with kids in the parent-AI co-reading mode?}} & The parent uses the auto-reading feature \\ \cline{2-2}
                                                                                                & The parent reads the story by themselves \\ \cline{2-2}
                                                                                                & The parent lets the child read the story \\ \hline
             \multirow{3}{*}{\parbox{5cm}{How did parents ask questions in the the parent-AI co-reading mode?}} & The parent asks a generated question themselves \\ \cline{2-2}

                                     & The parent uses the chatbot to ask a generated question  \\ \cline{2-2}
                                     
                                     & The parent asks a question of their own \\ \hline
             \multirow{3}{*}{\parbox{5cm}{How did parents use follow-up questions in the parent-AI co-reading mode?}} 
                                                                                    & The parent asks a provided follow-up question by themselves \\ \cline{2-2}

                                                                                    & The parent uses the chatbot to ask a follow-up question \\ \cline{2-2}
                                                                                    
                                                                                    & The parent asks a follow-up question of their own \\  \hline
             \multirow{2}{*}{\parbox{5cm}{How did parents configure the questions in the child-alone mode?}} & The parent uses the default setting \\ \cline{2-2}
                                                                        & The parent configures the preferred types of generated questions \\  \hline
             \multirow{3}{*}{\parbox{5cm}{How did children interact with the chatbot in the child-alone mode?}} & The child answers an initial question from the bot \\ \cline{2-2} & The child reattempts a question when the answer was incorrect \\ \cline{2-2}
                                                                        & The child answers a follow-up question from the bot \\ \hline
        \end{tabular}
    \centering
    \caption{The behaviors of interest from screen recordings} 
    \vspace{-0.4cm}
    \label{tab:behavior}
    \end{table*}

    \subsubsection{Observation of user behaviors}
    \label{sec:obeservation_result}
    To understand how parents and children used the StoryBuddy in the study, we analyzed the observed user behaviors from screen recordings in the study. We aim to answer the following questions: (1) How did parents read story with kids in the parent-AI co-reading mode? (2) How did parents ask questions in the parent-AI co-reading mode? (3) How did parents use follow-up questions in the parent-AI co-reading mode? (4) How did parents configure the questions in the child-alone mode? (5) How did children interact with the chatbot in the child-alone mode? 
    
    \paragraph{Analysis methods}
    Based on the above research questions, we came up a list of behaviors of interest for the annotation (Table~\ref{tab:behavior}). Note that those behaviors are rather objective (e.g. the user interacts with a certain feature in the system) with little space for subjective interpretation. One author carefully went through the screen recording videos of all study sessions to annotate these behaviors and count their occurrences.
    
    \paragraph{Findings}
    Parents used different reading strategies in the parent-AI co-reading mode. A factor that affected the reading strategy of choice was the age of their children. Five parents used the auto-reading feature throughout the study. Four parents who all have younger children (5 or younger) decided to read the story by themselves. Compared with the text-to-speech auto-reading, they deliberately made the reading more emotional and slower. On the contrary, three parents of the older children from 6--7 had their children lead the story reading and only helped them with new words. These findings confirmed the usefulness of supporting multiple reading strategies (and a mix of them) in Storybuddy. Such differences may be attributed to the differences between children in the pre-reading stage and the early-reading stage, where younger children rely on the sound structure of spoken language but older children can read stories more independently~\cite{norman1987stages}. It also indicates the opportunity for supporting better age-based customization, which we will discuss in Section~\ref{sec:future_work}.
    
    \zzhighlight{Parents used the generated questions in different ways. Two parents asked all the displayed questions in the given order on every page. Seven parents only selected some questions to ask, when we ask them about how they picked the questions to use in the post-study interview, they reported that their selections were based on their intuition of whether the questions were comprehensible to the kids. For the generated follow-up questions, five parents did not ask them at all throughout the reading. Four parents asked those questions sometimes. They tended to skip follow-up questions when they noticed the child became impatient, or when those questions were less relevant with previous ones. Lastly, two parents used the chatbot to ask follow-up questions.}
    
    \zzhighlight{For the question type configuration in the child-alone mode, six parents directly used the default settings without any modification. The other six parents reviewed and made some modifications to the preferred question types to be generated. This implies that some parents may not want to bother with the configuration, while others take advantage of the configuration option to better adapt the system to their goals and needs.}
    
    \zzhighlight{We also analyzed how children interacted with the chatbot in the child-alone mode. There was one child who simply used the StoryBuddy to read the book and skipped all interactions with the chatbot. All other eleven children interacted with the chatbot. However, the extent to which they used the chatbot varied. Six of them only tried to answer the first question throughout the study and would move on to the next page no matter whether their answer was considered correct by the chatbot. Among the rest, three children were willing to make another attempt when their answers were incorrect, but would skip to the next page if their reattempt was still not accepted. Only two children kept interacting with the chatbot on a question until the answer got accepted. These observations indicate future design opportunities in more effectively engaging with children in multi-turn conversations to provide them with helpful guidance (especially in case of partially correct answers) that assists them in refining their answers.}

    \subsubsection{Post-study interview}
    \label{sec:interview_findings}

    We ended each study session with a 10-minute semi-structured interview with the parent. Besides following up on their post-study questionnaire responses, we further asked the parent about the difficulties they encountered during the study, whether and how the StoryBuddy helped the children and parents on storytelling, how they would use the StoryBuddy in their daily life, and their suggestions on system improvement.
    
    \zzhighlight{ \paragraph{Analysis methods}
    In the guidance of established open coding methods~\cite{Braun2006UsingTA, Lazar2010ResearchMI}, two authors conducted a thematic analysis of interview transcripts to common themes with respect to user experiences, challenges, potential usage, and feedback. Specifically, each coder first individually went through and coded the transcripts of all the interview sessions using an inductive approach. For user quotes that did not include straightforward key terms, coders assigned researcher-denoted concepts as the code. Two coders discussed the themes emerging from the coding process and reached a consensus on the codebook. Then they independently mapped the extracted quotes to codes. The two coders reached a strong level of agreement in the interrater reliability with Cohen's Kappa $\kappa=0.81$.}

    \zzhighlight{ \paragraph{StoryBuddy as a useful language-learning tool for non-English speaking families}
    Four parents who were not native English speakers recognized the use of StoryBuddy for facilitating language learning through AI-facilitated story reading in non-English speaking families. The StoryBuddy system can teach children the ``\textit{correct pronunciation}'' (PB11) and ``\textit{vocabulary}'' (PB8) that parents were not good at. Besides, the StoryBuddy system allowed young children to enjoy storytelling even if the caregivers cannot speak English. For example, PB5 said ``\textit{I'm also thinking because we're bilingual family, and some of our family members, like grandparents, they cannot speak English. But if grandparents are caregivers, and they use this system, they can also still help their grandchild to do the English story reading.}'' These findings confirmed results from prior study on how conversational agents can support children's language development by serving as their language partners~\cite{xu_current:_2021}.}

    \zzhighlight{ \paragraph{Using StoryBuddy to develop and assess children's reading skills}
    Seven parents commented on StoryBuddy's value in developing and assessing children's reading skills. Specifically, they thought the generated questions for stories can ``\textit{cultivate children's critical reading}'' (PB1) and help them ``\textit{figure out what the key points were of the paragraph they were reading}'' (PB2).  Plus, one participant (PB4) believed StoryBuddy can engage his kid and make him more concentrated during the story-reading: ``\textit{if there's actually something like this that's able to help them (the children) be more engaging, they will be more concentrated on the story they are reading}'' (PB4). On the assessment side, the dashboard provides parents with ``\textit{a straightforward way to understand the children's current performance}'' (PB6) and ``\textit{know what aspects of reading skills their children need help on}'' (PB8). PB9 said, ``\textit{the dashboard is pretty good. It’s kind of capture what does the kid like or dislike (about question and story), and what they learn.}'' Such objectives of StoryBuddy reported by participants are consistent with prior literatures on how conversational agents can serve as effective partners that support the improvement of children's story comprehension~\cite{xu_same:_2021}.} 
    
    
    
    \zzhighlight{ \paragraph{Reducing parents' burdens in storytelling} Ten parents confirmed that StoryBuddy can reduce their burden by allowing flexible involvement and assisting them in coming up with questions, which was in the initial design goals (DS2 and DS6) and came from an original insight in the participatory design (Section~\ref{sec:conflict_parental_involvement}). StoryBuddy was found to be especially useful in time-conflict situations when ``\textit{they (children) want to read something but you (parents) are not available}'' (PB10), or be ``\textit{a supplementary tool when the parents feel tired or lazy}'' (PB3). However, PB3 also mentioned that the StoryBuddy cannot totally replace the parents' role in storytelling, because ``\textit{having a reading time together is a special time of enhancing parent-child relationship and those emotional parts cannot be offered by machine.}'' Nonetheless, parents agreed that the StoryBuddy would ``\textit{save their effort on coming up with questions}'' (PB5) and this embedded question-answering feature could ``\textit{make the storytelling more engaging}'' (PB4). }

    \zzhighlight{ \paragraph{Personalizing the reading experience}
    Nine parents commented on StoryBuddy's features for personalization, especially through its support of parent question configuration. They liked that they were able to control the types of generated questions, review generated questions and made changes as needed. Parents also pointed out several opportunities for further enhancing StoryBuddy's support of personalization. For example, PB10 advised that StoryBuddy can serve as a book recommender: ``\textit{since not every parent and children know the proper (difficulty) level, you can make this system recommend the books to kids, perhaps also based on their interests and reading history}'', though it may require the StoryBuddy to ``\textit{collect more book selections covering a wide range of topics}'' (PB10). Parents also suggested that the interaction flow of StoryBuddy could be more flexible to fit different family's story-reading practices.  For example, PB3 preferred to ``\textit{have the question-answering session at the end of the whole book instead of on every page}'' (PB3) and the current design could ``\textit{break the natural flow of parent-child storytelling}'' (PB3). While another parent really liked the questions-at-every-page approach: ``\textit{I liked the way that the questions being asked because it happened on every page. There are actually similar applications that read the whole story and ask questions at the end. In this way, the kids could easily forget the content and it is harder for them to answer the question (Her kid was nodding). I think this (the in-place question-asking) is the best feature you did in your app}''. These findings confirmed the importance of enabling the system to adapt to the unique contexts, preferences, and needs of parents and children as well as the established family practice in AI-enabled systems that facilitate parent-child interactions.}
    
    \zzhighlight{\paragraph{Opportunities for adapting StoryBuddy to different age groups}
    The interviews revealed opportunities for further personalizing the reading experience offered by StoryBuddy based on the age and the developmental stage of children. Although the dialogic reading approach is generally effective for children aged three to eight, which is the age group that StoryBuddy targets, it would be useful to further customize the reading experience for children of different ages within this group. For some older children, some generated questions could be ``\textit{too stupid and meaningless so that the kids lost their interest in using this system}'' (PB7), while the younger ones found a small number of questions were ``\textit{very hard and beyond understanding}'' (PB8). The age-based personalization can also apply to interaction strategies. The bot-guided reading approach in the parent absent mode generally worked quite well for younger children in our study, but the guidance may be ``too slow'' or even annoying for older children or children who are advanced in the development of reading skills, as PB12 said ``\textit{I think it may be useful for those younger kids, my son is able to read stories on his own, he may not need such a system to assist his reading.}'' Such differences correspond to the distinctions between children in the pre-reading stage and those in the early-reading stage, where older children start to learn how to read independently without relying on the sound structures of spoken language~\cite{norman1987stages, hoien1988stages}.}

    \zzhighlight{\paragraph{Usability issues}
    The general attitude towards the usability of StoryBuddy was positive. However, the participants still experienced some  specific usability issues during the use. The first problem was the inaccurate speech recognition in the Google Cloud Speech Recognition API. It sometimes misinterpreted the children's speech or stopped the recognition early. As PB9 said ``\textit{the children would not be frustrated because of the incorrect recognition of their answers}''. Parents also found the size of some widgets in StoryBuddy were too small for their kids: ``\textit{the buttons were too small for children to use, especially when they use fingers on iPad}'' (PB9). Besides, parents also suggested to ``\textit{make the interface more colorful or add more cartoon element}'' (PB3) so as to engage kids to use.}

    \subsection{Threats to Validity}
    \label{sec:threats_to_validity}
    
    A potential threat to the validity of the user study results is the sample bias in our participants. Since our participant recruitment was done in a University community, all the participating parents had at least a bachelor's degree. More than half of the participating parents had received or were pursuing a graduate degree. Parents in underrepresented racial groups, from lower socioeconomic backgrounds, or with non-traditional family structures were also underrepresented in our participants. 
    
    Another threat lies at the ecological validity---The study was done remotely through video calls and screen-sharing via Zoom. The setting of the study did not closely resemble the realistic setting where StoryBuddy might be used. For example, when we tested the automated bot-reading mode of StoryBuddy, the parent was asked to refrain from intervening the child-agent interaction. However, the parent was still co-located and within the sight of the child. In realistic usage of this mode, the parent will likely be completely absent.
    
    It would be also interesting to investigate how much of a role did the \textit{novelty factor} of interacting with a conversational agent played in the strong child engagement during the storytelling sessions in our user study and to measure how the effect of such novelty factor would change over time. 
    
    We plan to address these potential threats to validity in the future through (1) first, a larger-scale field deployment study with a more representative user population; and (2) eventually, a public release of StoryBuddy to the general public (detail in Section~\ref{sec:future_work}).

\section{Discussion and Design Implications}

The results from our user study suggest that parents can successfully collaborate with StoryBuddy to create interactive storytelling experiences for their children. StoryBuddy also performed well in keeping the child engaged and entertained. Its interaction design allowed changes in how the child interacts with the parent and the agent throughout the storytelling process so that the child did not easily get bored.

Below, we discuss the implications and design themes emerged from our work~\cite{hook_2012_strong, Barendregt_intermediate_2017}. 
\subsection{Partial Automation in Parent-AI Collaboration}

From the lens of mixed-initiative interaction~\cite{horivitz1999principles} and human-AI collaboration~\cite{wang_human_2020,wang_designing_2021}, a central issue to consider in designing parent-AI collaborative systems is to identify opportunities for \textit{partial automation} based on the capabilities of the AI system and the capabilities of the parent~\cite{lau_why_2009}. For example, in StoryBuddy, the key AI capability we leveraged is that its back-end model can quickly generate questions, answers, and possible follow-up questions of seven types from the textual content of \textit{any} story. However, it is important to recognize that the parent possesses knowledge about the task and the context that is unknown to the system. For instance, they know the preferences of their child, their own goals for child skill development, and the constraints in the current context. Such knowledge allows the parent to customize the generated questions accordingly.   

To accommodate partial automation, the interaction flow of the system needs to cope with interruptions and resumptions. In the opposite direction, the parent also needs to be attentive and ready to contribute when needed. For example, in the parent-AI co-reading mode of StoryBuddy, the parent needs to make on-the-fly decisions on whether to use an AI-recommended question or an AI-generated follow-up question in synchronous storytelling. However, since the parent is also the one telling the story in this mode, they are able to adjust the pace of the parent-child interaction. If we had designed a system where the system took initiative in telling the story but the parent needed to make decisions on what questions to ask and how to ask them while keeping up with the pace of the system, coordinating harmonious collaboration between the AI system and the parent would be much more challenging.

\subsection{Supporting Flexibility in Parent Involvement}
Beyond accounting for the capabilities of the parent and the AI system when designing partial automation, we also need to consider the parent's varied preferences and changing availability in their involvement. For example, in our formative study, PD process, and user study, we encountered parents who cared deeply about the educational goals and skill assessment features in StoryBuddy. They wished to have fine granularity control on the types of questions generated by the system and the specific follow-up questions to use. For them, we designed the question preference configuration panel (Section~\ref{sec:question_preference}), the interactive dashboard for child progress tracking and performance assessment (Section~\ref{sec:dashboard}), and the features for manually editing questions and answers when configuring the StoryBuddy agent in the automated bot-reading mode. However, the use of all these features is optional---a parent who mainly uses StoryBuddy for engagement and entertainment goals could skip all these steps if they wish so.

Another design consideration for parent involvement is how we can help parents stay involved when they are unavailable for synchronous parent-child interactions. The conflict between (1) parents' desire for fulfilling their children's storytelling needs and staying involved in the process; and (2) parents' limited and constrained time availability was a recurring theme in our studies. This problem is aggravated by the increasingly common work-from-home arrangement for parents. In work-from-home situations, parents may \textit{seem available} to children for interaction since they are physically home when they are, in fact, unavailable. An interactive storytelling agent that can keep children engaged without requiring parental intervention would be particularly useful in such scenarios. To address this issue, the core strategy that StoryBuddy used is to turn synchronous involvement into asynchronous involvement---While the delivery of the story content and the interaction for question-answering is facilitated by the StoryBuddy agent, the parent has (1) control over the content of the interaction through the configuration \textit{before} a storytelling session; and (2) knowledge on the progress and the performance of the child through viewing the dashboard \textit{after} a storytelling session.


\subsection{\tlhighlight{Role of AI in Parent-Child Interaction}}

\tlhighlight{The design of StoryBuddy introduced an AI system into an \textit{existing} activity that previously involved only the parent and the child: parent-child joint reading, as a result, an important challenge is to determine the appropriate role that the AI should play. Should the AI system be an assistant, a peer, a companion, an agent of the parent, or something else? To complicate the problem, the activity of parent-child joint reading usually fulfills a combination of multiple goals: relationship building, skill development and assessment, and entertainment. Therefore, it is crucial to consider the roles that the AI system would play in each part of the activity and how its roles contribute (positively or negatively) to the user's goals.}

For example, in our design, StoryBuddy assists the parent by helping them identify opportunities for asking questions, recommending questions to use, and proposing follow-up questions. \tlhighlight{These forms of assistance help parents come up with better questions for fulfilling their skill development and assessment goal, reduce their cognitive load so that they can allocate more attention to the interaction with their child for the relationship-building goal, and keep their child engaged and entertained through the occasional interaction between the child and the agent. However, all these forms of AI involvement do not lessen the central role of the parent in their interactions with the child. In its chatbot form, StoryBuddy starts to act as a companion or a peer to children as it communicates with them through natural language dialogs---the agent may also be leveraged as a third-party mediator that facilitates parent-child communication as reported in~\cite{beneteau_parenting:_2020}.}

There are still issues with the current design of StoryBuddy and opportunities for how its roles can be refined. For example, as discussed in Section~\ref{sec:interview_findings}, a parent criticized that StoryBuddy interfered with their normal story reading flow by proposing to ask questions after every section instead of at the end of the whole story (which they often do in their parent-child joint reading practice without AI involvement). Parents also suggested that StoryBuddy may further help with the entertainment and engagement goal by combining its tracking of child progress with gamification. For example, it can give out virtual ``badges'' when the child reaches certain achievements and milestones.

In future projects on designing AI systems for parents and children, it is crucial to first start with formative studies that uncover the multi-faceted goals of the parent and the child. After that, multiple design iterations with intensive user participation in the user-centered design process are needed in order to carefully define the appropriate roles that the AI system should play.

\section{Future Work}
\label{sec:future_work}

There are several directions for planned future work. The current version of StoryBuddy focuses on parent-led and agent-led question-answering. A different variation of question-answering that we plan to support in the future is child-led question-answering, where the child asks questions about the story plot and the agent can answer these questions and ask appropriate follow-up questions. A parent in our PD session reported currently using this strategy in their parent-child joint story-reading (without the use of digital assistance) with success. Prior literature~\cite{tamura2017effects} suggested that a bot that can answer children's questions during the storytelling process is preferred by children. We may also explore the design space of other interaction approaches in interactive storytelling, such as reflective storytelling~\cite{hubbard2021chid}. Another useful enhancement on StoryBuddy's question generation capabilities is to support image-based questions about the content of the visual illustrations in the storybook in addition to questions about the textual contents. This idea came up in our discussion with multiple parents in both the PD process and the usability study.

While the current version of StoryBuddy supports a wide range of device types including smartphones, tablets, laptops, and desktops, another type of device that would be useful to support is the smart speaker. Some parents expressed their concerns about limiting the ``screen time'' for their children in our studies. The child may also, for example, switch away from StoryBuddy and play video games instead when they use StoryBuddy on a smartphone or a tablet when the parent is absent. Supporting running StoryBuddy on smart speakers such as Amazon Echo, Apple HomePod, or Google Home can alleviate these concerns. However, adapting StoryBuddy to a screenless smart speaker raises additional design challenges in, for example, maintaining conversational context in longer multi-turn conversations, facilitating turn-taking in question-answering, grounding questions back to the story content, and displaying the progress of storytelling sessions. We plan to investigate these issues for future work.

A direction of future work informed by participatory design and user study results is to make the back-end question generation model adaptive to parent preferences and child interactions. With the current version of StoryBuddy, the parent can customize the questions through (1) configuring the back-end question generation model on the preferred question types (Section~\ref{sec:question_preference}); and (2) selecting and editing the generated questions for each story (Figure~\ref{fig:parentPreconfiguration}). However, this process can get tedious if the parent wishes to go through all the questions. Once the configuration is done, the agent's question-asking plan also remains static without the ability to adjust based on child interactions. To address these limitations, we plan to explore inserting a model in StoryBuddy that learns the parent's preferences while they select and tweak the questions to use. After the parent finishes configuring a few pages, the model can automatically adjust the generating questions for the rest of the story according to the learned parent preferences to reduce parent effort. At the story-telling time, another model can track child performance and engagement in order to automatically adjust the difficulty and the types of questions in real-time.  

\tlhighlight{Another direction for future study is to understand how children from different age groups interact with StoryBuddy differently and how the design of StoryBuddy can be extended to accommodate their diverse needs. As reported in the results from the post-study interviews in Section~\ref{sec:interview_findings}, parents identified several opportunities in the current version of StoryBuddy to better accommodate the needs of children of different ages: on the question-generation back-end, the model could adjust the complexity, the vocabulary used, and the cognitive skill required in its generated questions based on the development progress of the child. The interactive strategy on e.g., how much guidance to provide during story reading in the parent absent mode can also vary according to the needs of children of different ages. Prior literature also suggested that children from the pre-reading and early-reading groups enjoy and benefit from graphical contents (e.g., pictures) in storybooks as much as textual contents~\cite{donald1983use, newton1995role}. While the current version of StoryBuddy displays pictures from the original book in its story reading interface, none of its interactions with parents and children refers to the content of the pictures. For future work, it would be useful to investigate ways to generate questions about pictures in the storybook or questions that connect the textual contents of a storybook with its graphical contents.} 

\yxhighlight{In addition, our user study focused on using StoryBuddy in the home settings given the important role home literacy environments play in children's literacy development. We are also interested in exploring how automatic question generation can be used to support teachers in classroom instructions. It is conceivable that our system has the potential to enable more personalized interactive reading instruction. First, teachers can use our system to easily generate customized reading resources for students based on student needs. Second, teachers can better monitor students' reading comprehension by tracking the students' performance in answering dialogic questions in real-time. This will allow the early intervention of students who may be at risk for reading difficulty. }

\tlhighlight{Another currently under-explored design space in StoryBuddy is the child-led intervention. The current support interactions in StoryBuddy are mostly driven by parents---parents prepare the co-reading companion by configuring the question types and selecting the mode to use. In the parent-child co-reading mode, the parent is the one who controls the flow of the co-reading and decides when to ask a question and what question to ask. The child-led approach, as described in~\cite{slovak_child_2018}, has the potential to directly engage with children and empower children to learn by themselves rather than relying on adults. In order to support this in future versions of StoryBuddy, we need to improve the design of configuration interfaces with the capability and the preferences of children in mind so that they can configure the StoryBuddy agent themselves. Children may also use assistance in e.g., recommending a potential storybook of interest. Future design activities are also needed to explore how to foster relationships and rapport-building between children and the AI-enabled story-reading companion in a child-led rather than parent-driven fashion. To take another step forward, another direction to further support child-led interaction is to involve children in the creation and development of stories. Such collective storytelling among children, AI systems, and (optionally) parents can give children structure and significance to the world around them, facilitating pedagogical and psychological development~\cite{Barendregt_intermediate_2017}}


Lastly, we are hoping to deploy StoryBuddy with a larger group of users and eventually release the system for public use. Although the design of StoryBuddy was informed by formative study and participatory design results, and the usability of StoryBuddy has been validated by a user study, we hope to further understand how parents configure different kinds of stories for children, how parents choose between the two modes, how parents use the different interaction paradigms in the parent-AI co-reading mode, and how useful StoryBuddy is for users in realistic contexts. The main goal of the deployment would be to study StoryBuddy within its intended context of use. Another goal is to study the use of StoryBuddy in a more representative user population. As discussed in Section~\ref{sec:threats_to_validity}, there are demographic biases in the participant population of our user study. We seek to recruit a more diverse group of users in the deployment and public release of StoryBuddy. 

\section{Conclusion}
This paper presented StoryBuddy, a new system that allowed parents to collaborate with an AI system in creating storytelling experiences with interactive questioning-answering. Informed by results from a formative study and a participatory design study, we designed two distinct modes and several dynamic interaction paradigms that supported flexible degrees of parent involvement in (1) the configuration of the interactive storytelling experience before the storytelling; (2) the parent-AI co-delivery of the story content to the child during the storytelling; and (3) the tracking and assessment of child progress and performance after the storytelling. A user study with 12 pairs of parents and children found that StoryBuddy was effective in providing parents with desired levels of control and involvement while maintaining children's engagement in the storytelling process. Parents and children considered StoryBuddy useful, helpful, and likable.  

\begin{acks}
This work was supported in part by an Asia Research Collaboration Grant from Notre Dame International, a Google Cloud Research Credit Grant, the Rensselaer-IBM AI Research Collaboration (\url{http://airc.rpi.edu}) part of the IBM AI Horizons Network (\url{http://ibm.biz/AIHorizons}), and the National Science Foundation (Grant No. 1906321 and 2115382). The FairytaleQA dataset used in this project is funded by Schmidt Futures. Any opinions, findings or recommendations expressed here are those of the authors and do not necessarily reflect views of the sponsors. 

We would like to thank our anonymous reviewers for their feedback and study participants for their participation of our studies. We are grateful to Mark Warschauer, Yuwen Lu, Zheng Ning, and Yaxing Yao for useful discussions, Tiffany Iong for illustrating the storyboards, Alondra Perez for helping conduct the user studies, and research assistants at UC Irvine for assisting with the study coordination and preparing the dataset. 
\end{acks}

\balance
\bibliographystyle{ACM-Reference-Format}
\bibliography{references}


\begin{thebibliography}{89}


\ifx \showCODEN    \undefined \def \showCODEN     #1{\unskip}     \fi
\ifx \showDOI      \undefined \def \showDOI       #1{#1}\fi
\ifx \showISBNx    \undefined \def \showISBNx     #1{\unskip}     \fi
\ifx \showISBNxiii \undefined \def \showISBNxiii  #1{\unskip}     \fi
\ifx \showISSN     \undefined \def \showISSN      #1{\unskip}     \fi
\ifx \showLCCN     \undefined \def \showLCCN      #1{\unskip}     \fi
\ifx \shownote     \undefined \def \shownote      #1{#1}          \fi
\ifx \showarticletitle \undefined \def \showarticletitle #1{#1}   \fi
\ifx \showURL      \undefined \def \showURL       {\relax}        \fi
\providecommand\bibfield[2]{#2}
\providecommand\bibinfo[2]{#2}
\providecommand\natexlab[1]{#1}
\providecommand\showeprint[2][]{arXiv:#2}

\bibitem[\protect\citeauthoryear{Alexandrakis, Chorianopoulos, and
  Tselios}{Alexandrakis et~al\mbox{.}}{2019}]%
        {Alexandrakis2019Insights}
\bibfield{author}{\bibinfo{person}{Diogenis Alexandrakis},
  \bibinfo{person}{Konstantinos Chorianopoulos}, {and}
  \bibinfo{person}{Nikolaos Tselios}.} \bibinfo{year}{2019}\natexlab{}.
\newblock \showarticletitle{Insights on Older Adults' Attitudes and Behavior
  Through the Participatory Design of an Online Storytelling Platform}. In
  \bibinfo{booktitle}{\emph{Human-Computer Interaction -- INTERACT 2019}},
  \bibfield{editor}{\bibinfo{person}{David Lamas}, \bibinfo{person}{Fernando
  Loizides}, \bibinfo{person}{Lennart Nacke}, \bibinfo{person}{Helen Petrie},
  \bibinfo{person}{Marco Winckler}, {and} \bibinfo{person}{Panayiotis
  Zaphiris}} (Eds.). \bibinfo{publisher}{Springer International Publishing},
  \bibinfo{address}{Cham}, \bibinfo{pages}{465--474}.
\newblock
\showISBNx{978-3-030-29381-9}


\bibitem[\protect\citeauthoryear{Ashktorab, Jain, Liao, and Weisz}{Ashktorab
  et~al\mbox{.}}{2019}]%
        {ashktorab2019resilient}
\bibfield{author}{\bibinfo{person}{Zahra Ashktorab}, \bibinfo{person}{Mohit
  Jain}, \bibinfo{person}{Q~Vera Liao}, {and} \bibinfo{person}{Justin~D
  Weisz}.} \bibinfo{year}{2019}\natexlab{}.
\newblock \showarticletitle{Resilient Chatbots: Repair Strategy Preferences for
  Conversational Breakdowns}. In \bibinfo{booktitle}{\emph{Proceedings of the
  2019 CHI Conference on Human Factors in Computing Systems}}. ACM,
  \bibinfo{pages}{254}.
\newblock


\bibitem[\protect\citeauthoryear{Barendregt, Torgersson, Eriksson, and
  B\"{o}rjesson}{Barendregt et~al\mbox{.}}{2017}]%
        {Barendregt_intermediate_2017}
\bibfield{author}{\bibinfo{person}{Wolmet Barendregt}, \bibinfo{person}{Olof
  Torgersson}, \bibinfo{person}{Eva Eriksson}, {and} \bibinfo{person}{Peter
  B\"{o}rjesson}.} \bibinfo{year}{2017}\natexlab{}.
\newblock \showarticletitle{Intermediate-Level Knowledge in Child-Computer
  Interaction: A Call for Action}. In \bibinfo{booktitle}{\emph{Proceedings of
  the 2017 Conference on Interaction Design and Children}} (Stanford,
  California, USA) \emph{(\bibinfo{series}{IDC '17})}.
  \bibinfo{publisher}{Association for Computing Machinery},
  \bibinfo{address}{New York, NY, USA}, \bibinfo{pages}{7–16}.
\newblock
\showISBNx{9781450349215}
\urldef\tempurl%
\url{https://doi.org/10.1145/3078072.3079719}
\showDOI{\tempurl}


\bibitem[\protect\citeauthoryear{Beneteau, Boone, Wu, Kientz, Yip, and
  Hiniker}{Beneteau et~al\mbox{.}}{2020}]%
        {beneteau_parenting:_2020}
\bibfield{author}{\bibinfo{person}{Erin Beneteau}, \bibinfo{person}{Ashley
  Boone}, \bibinfo{person}{Yuxing Wu}, \bibinfo{person}{Julie~A. Kientz},
  \bibinfo{person}{Jason Yip}, {and} \bibinfo{person}{Alexis Hiniker}.}
  \bibinfo{year}{2020}\natexlab{}.
\newblock \bibinfo{booktitle}{\emph{Parenting with Alexa: Exploring the
  Introduction of Smart Speakers on Family Dynamics}}.
\newblock \bibinfo{publisher}{Association for Computing Machinery},
  \bibinfo{address}{New York, NY, USA}, \bibinfo{pages}{1–13}.
\newblock
\showISBNx{9781450367080}
\urldef\tempurl%
\url{https://doi.org/10.1145/3313831.3376344}
\showURL{%
\tempurl}


\bibitem[\protect\citeauthoryear{Bentley, Luvogt, Silverman, Wirasinghe, White,
  and Lottridge}{Bentley et~al\mbox{.}}{2018}]%
        {bentley2018understanding}
\bibfield{author}{\bibinfo{person}{Frank Bentley}, \bibinfo{person}{Chris
  Luvogt}, \bibinfo{person}{Max Silverman}, \bibinfo{person}{Rushani
  Wirasinghe}, \bibinfo{person}{Brooke White}, {and} \bibinfo{person}{Danielle
  Lottridge}.} \bibinfo{year}{2018}\natexlab{}.
\newblock \showarticletitle{Understanding the Long-Term Use of Smart Speaker
  Assistants}.
\newblock \bibinfo{journal}{\emph{Proc. ACM Interact. Mob. Wearable Ubiquitous
  Technol.}} \bibinfo{volume}{2}, \bibinfo{number}{3}, Article
  \bibinfo{articleno}{91} (\bibinfo{date}{Sept.} \bibinfo{year}{2018}),
  \bibinfo{numpages}{24}~pages.
\newblock
\urldef\tempurl%
\url{https://doi.org/10.1145/3264901}
\showDOI{\tempurl}


\bibitem[\protect\citeauthoryear{Bhatti, Stelter, and McCrickard}{Bhatti
  et~al\mbox{.}}{2021}]%
        {bhatti2021conversational}
\bibfield{author}{\bibinfo{person}{Neelma Bhatti}, \bibinfo{person}{Timothy~L
  Stelter}, {and} \bibinfo{person}{D~Scott McCrickard}.}
  \bibinfo{year}{2021}\natexlab{}.
\newblock \showarticletitle{Conversational User Interfaces As Assistive
  interlocutors For Young Children's Bilingual Language Acquisition}.
\newblock \bibinfo{journal}{\emph{arXiv preprint arXiv:2103.09228}}
  (\bibinfo{year}{2021}).
\newblock


\bibitem[\protect\citeauthoryear{Birks, Chapman, and Francis}{Birks
  et~al\mbox{.}}{2008}]%
        {birks2008memoing}
\bibfield{author}{\bibinfo{person}{Melanie Birks}, \bibinfo{person}{Ysanne
  Chapman}, {and} \bibinfo{person}{Karen Francis}.}
  \bibinfo{year}{2008}\natexlab{}.
\newblock \showarticletitle{Memoing in qualitative research: Probing data and
  processes}.
\newblock \bibinfo{journal}{\emph{Journal of research in nursing}}
  \bibinfo{volume}{13}, \bibinfo{number}{1} (\bibinfo{year}{2008}),
  \bibinfo{pages}{68--75}.
\newblock


\bibitem[\protect\citeauthoryear{Blewitt, Rump, Shealy, and Cook}{Blewitt
  et~al\mbox{.}}{2009}]%
        {blewitt2009shared}
\bibfield{author}{\bibinfo{person}{Pamela Blewitt}, \bibinfo{person}{Keiran~M
  Rump}, \bibinfo{person}{Stephanie~E Shealy}, {and}
  \bibinfo{person}{Samantha~A Cook}.} \bibinfo{year}{2009}\natexlab{}.
\newblock \showarticletitle{Shared book reading: When and how questions affect
  young children's word learning.}
\newblock \bibinfo{journal}{\emph{Journal of Educational Psychology}}
  \bibinfo{volume}{101}, \bibinfo{number}{2} (\bibinfo{year}{2009}),
  \bibinfo{pages}{294}.
\newblock


\bibitem[\protect\citeauthoryear{Bodker}{Bodker}{1999}]%
        {bodker1999scenarios}
\bibfield{author}{\bibinfo{person}{Susanne Bodker}.}
  \bibinfo{year}{1999}\natexlab{}.
\newblock \showarticletitle{Scenarios in user-centred design-setting the stage
  for reflection and action}. In \bibinfo{booktitle}{\emph{Proceedings of the
  32nd Annual Hawaii International Conference on Systems Sciences. 1999.
  HICSS-32. Abstracts and CD-ROM of Full Papers}}. IEEE,
  \bibinfo{pages}{11--pp}.
\newblock


\bibitem[\protect\citeauthoryear{Braun and Clarke}{Braun and Clarke}{2006}]%
        {Braun2006UsingTA}
\bibfield{author}{\bibinfo{person}{Virginia Braun} {and}
  \bibinfo{person}{Victoria Clarke}.} \bibinfo{year}{2006}\natexlab{}.
\newblock \showarticletitle{Using thematic analysis in psychology}.
\newblock \bibinfo{journal}{\emph{Qualitative Research in Psychology}}
  \bibinfo{volume}{3} (\bibinfo{year}{2006}), \bibinfo{pages}{101 -- 77}.
\newblock


\bibitem[\protect\citeauthoryear{Cowan, Pantidi, Coyle, Morrissey, Clarke,
  Al-Shehri, Earley, and Bandeira}{Cowan et~al\mbox{.}}{2017}]%
        {Cowan:2017:IHY:3098279.3098539}
\bibfield{author}{\bibinfo{person}{Benjamin~R. Cowan}, \bibinfo{person}{Nadia
  Pantidi}, \bibinfo{person}{David Coyle}, \bibinfo{person}{Kellie Morrissey},
  \bibinfo{person}{Peter Clarke}, \bibinfo{person}{Sara Al-Shehri},
  \bibinfo{person}{David Earley}, {and} \bibinfo{person}{Natasha Bandeira}.}
  \bibinfo{year}{2017}\natexlab{}.
\newblock \showarticletitle{"What Can I Help You with?": Infrequent Users'
  Experiences of Intelligent Personal Assistants}. In
  \bibinfo{booktitle}{\emph{Proceedings of the 19th International Conference on
  Human-Computer Interaction with Mobile Devices and Services}} (Vienna,
  Austria) \emph{(\bibinfo{series}{MobileHCI '17})}. \bibinfo{publisher}{ACM},
  \bibinfo{address}{New York, NY, USA}, Article \bibinfo{articleno}{43},
  \bibinfo{numpages}{12}~pages.
\newblock
\showISBNx{978-1-4503-5075-4}
\urldef\tempurl%
\url{https://doi.org/10.1145/3098279.3098539}
\showDOI{\tempurl}


\bibitem[\protect\citeauthoryear{Dietz, Le, Tamer, Han, Gweon, Murnane, and
  Landay}{Dietz et~al\mbox{.}}{2021}]%
        {dietz_storycoder_2021}
\bibfield{author}{\bibinfo{person}{Griffin Dietz}, \bibinfo{person}{Jimmy~K
  Le}, \bibinfo{person}{Nadin Tamer}, \bibinfo{person}{Jenny Han},
  \bibinfo{person}{Hyowon Gweon}, \bibinfo{person}{Elizabeth~L Murnane}, {and}
  \bibinfo{person}{James~A. Landay}.} \bibinfo{year}{2021}\natexlab{}.
\newblock \bibinfo{booktitle}{\emph{StoryCoder: Teaching Computational Thinking
  Concepts Through Storytelling in a Voice-Guided App for Children}}.
\newblock \bibinfo{publisher}{Association for Computing Machinery},
  \bibinfo{address}{New York, NY, USA}.
\newblock
\showISBNx{9781450380966}
\urldef\tempurl%
\url{https://doi.org/10.1145/3411764.3445039}
\showURL{%
\tempurl}


\bibitem[\protect\citeauthoryear{Donald}{Donald}{1983}]%
        {donald1983use}
\bibfield{author}{\bibinfo{person}{DR Donald}.}
  \bibinfo{year}{1983}\natexlab{}.
\newblock \showarticletitle{The use and value of illustrations as contextual
  information for readers at different progress and developmental levels}.
\newblock \bibinfo{journal}{\emph{British Journal of Educational Psychology}}
  \bibinfo{volume}{53}, \bibinfo{number}{2} (\bibinfo{year}{1983}),
  \bibinfo{pages}{175--185}.
\newblock


\bibitem[\protect\citeauthoryear{Dong, Yang, Wang, Wei, Liu, Wang, Gao, Zhou,
  and Hon}{Dong et~al\mbox{.}}{2019}]%
        {dong2019unified}
\bibfield{author}{\bibinfo{person}{Li Dong}, \bibinfo{person}{Nan Yang},
  \bibinfo{person}{Wenhui Wang}, \bibinfo{person}{Furu Wei},
  \bibinfo{person}{Xiaodong Liu}, \bibinfo{person}{Yu Wang},
  \bibinfo{person}{Jianfeng Gao}, \bibinfo{person}{Ming Zhou}, {and}
  \bibinfo{person}{Hsiao-Wuen Hon}.} \bibinfo{year}{2019}\natexlab{}.
\newblock \showarticletitle{Unified language model pre-training for natural
  language understanding and generation}.
\newblock \bibinfo{journal}{\emph{arXiv preprint arXiv:1905.03197}}
  (\bibinfo{year}{2019}).
\newblock


\bibitem[\protect\citeauthoryear{Du, Shao, and Cardie}{Du
  et~al\mbox{.}}{2017}]%
        {du-etal-2017-learning}
\bibfield{author}{\bibinfo{person}{Xinya Du}, \bibinfo{person}{Junru Shao},
  {and} \bibinfo{person}{Claire Cardie}.} \bibinfo{year}{2017}\natexlab{}.
\newblock \showarticletitle{Learning to Ask: Neural Question Generation for
  Reading Comprehension}. In \bibinfo{booktitle}{\emph{Proceedings of the 55th
  Annual Meeting of the Association for Computational Linguistics (Volume 1:
  Long Papers)}}. \bibinfo{publisher}{ACL}, \bibinfo{address}{Vancouver,
  Canada}, \bibinfo{pages}{1342--1352}.
\newblock
\urldef\tempurl%
\url{https://doi.org/10.18653/v1/P17-1123}
\showDOI{\tempurl}


\bibitem[\protect\citeauthoryear{Du, Zhang, Ramabadran, and Liu}{Du
  et~al\mbox{.}}{2021}]%
        {du_alexa:_2021}
\bibfield{author}{\bibinfo{person}{Yao Du}, \bibinfo{person}{Kerri Zhang},
  \bibinfo{person}{Sruthi Ramabadran}, {and} \bibinfo{person}{Yusa Liu}.}
  \bibinfo{year}{2021}\natexlab{}.
\newblock \bibinfo{booktitle}{\emph{“Alexa, What is That Sound?” A Video
  Analysis of Child-Agent Communication From Two Amazon Alexa Games}}.
\newblock \bibinfo{publisher}{Association for Computing Machinery},
  \bibinfo{address}{New York, NY, USA}, \bibinfo{pages}{513–520}.
\newblock
\showISBNx{9781450384520}
\urldef\tempurl%
\url{https://doi.org/10.1145/3459990.3465195}
\showURL{%
\tempurl}


\bibitem[\protect\citeauthoryear{Flack, Field, and Horst}{Flack
  et~al\mbox{.}}{2018}]%
        {flack2018effects}
\bibfield{author}{\bibinfo{person}{Zoe~M Flack}, \bibinfo{person}{Andy~P
  Field}, {and} \bibinfo{person}{Jessica~S Horst}.}
  \bibinfo{year}{2018}\natexlab{}.
\newblock \showarticletitle{The effects of shared storybook reading on word
  learning: A meta-analysis.}
\newblock \bibinfo{journal}{\emph{Developmental psychology}}
  \bibinfo{volume}{54}, \bibinfo{number}{7} (\bibinfo{year}{2018}),
  \bibinfo{pages}{1334}.
\newblock


\bibitem[\protect\citeauthoryear{Frohlich, Rachovides, Riga, Bhat, Frank,
  Edirisinghe, Wickramanayaka, Jones, and Harwood}{Frohlich
  et~al\mbox{.}}{2009}]%
        {frohlich_2019_storybank}
\bibfield{author}{\bibinfo{person}{David~M. Frohlich}, \bibinfo{person}{Dorothy
  Rachovides}, \bibinfo{person}{Kiriaki Riga}, \bibinfo{person}{Ramnath Bhat},
  \bibinfo{person}{Maxine Frank}, \bibinfo{person}{Eran Edirisinghe},
  \bibinfo{person}{Dhammike Wickramanayaka}, \bibinfo{person}{Matt Jones},
  {and} \bibinfo{person}{Will Harwood}.} \bibinfo{year}{2009}\natexlab{}.
\newblock \bibinfo{booktitle}{\emph{StoryBank: Mobile Digital Storytelling in a
  Development Context}}.
\newblock \bibinfo{publisher}{Association for Computing Machinery},
  \bibinfo{address}{New York, NY, USA}, \bibinfo{pages}{1761–1770}.
\newblock
\showISBNx{9781605582467}
\urldef\tempurl%
\url{https://doi.org/10.1145/1518701.1518972}
\showURL{%
\tempurl}


\bibitem[\protect\citeauthoryear{Frude and Killick}{Frude and Killick}{2011}]%
        {frude2011family}
\bibfield{author}{\bibinfo{person}{Neil Frude} {and} \bibinfo{person}{Steve
  Killick}.} \bibinfo{year}{2011}\natexlab{}.
\newblock \showarticletitle{Family storytelling and the attachment
  relationship}.
\newblock \bibinfo{journal}{\emph{Psychodynamic Practice}}
  \bibinfo{volume}{17}, \bibinfo{number}{4} (\bibinfo{year}{2011}),
  \bibinfo{pages}{441--455}.
\newblock


\bibitem[\protect\citeauthoryear{Garg and Sengupta}{Garg and Sengupta}{2020a}]%
        {garg2020conversational}
\bibfield{author}{\bibinfo{person}{Radhika Garg} {and}
  \bibinfo{person}{Subhasree Sengupta}.} \bibinfo{year}{2020}\natexlab{a}.
\newblock \showarticletitle{Conversational Technologies for In-home Learning:
  Using Co-Design to Understand Children's and Parents' Perspectives}. In
  \bibinfo{booktitle}{\emph{Proceedings of the 2020 CHI conference on human
  factors in computing systems}}. \bibinfo{pages}{1--13}.
\newblock


\bibitem[\protect\citeauthoryear{Garg and Sengupta}{Garg and Sengupta}{2020b}]%
        {garg_conversational:_2020}
\bibfield{author}{\bibinfo{person}{Radhika Garg} {and}
  \bibinfo{person}{Subhasree Sengupta}.} \bibinfo{year}{2020}\natexlab{b}.
\newblock \bibinfo{booktitle}{\emph{Conversational Technologies for In-Home
  Learning: Using Co-Design to Understand Children's and Parents'
  Perspectives}}.
\newblock \bibinfo{publisher}{Association for Computing Machinery},
  \bibinfo{address}{New York, NY, USA}, \bibinfo{pages}{1–13}.
\newblock
\showISBNx{9781450367080}
\urldef\tempurl%
\url{https://doi.org/10.1145/3313831.3376631}
\showURL{%
\tempurl}


\bibitem[\protect\citeauthoryear{Grudin and Pruitt}{Grudin and Pruitt}{2002}]%
        {grudin2002ppd}
\bibfield{author}{\bibinfo{person}{Jonathan Grudin} {and} \bibinfo{person}{John
  Pruitt}.} \bibinfo{year}{2002}\natexlab{}.
\newblock \showarticletitle{Personas, Participatory Design and Product
  Development: An Infrastructure for Engagement}. In
  \bibinfo{booktitle}{\emph{Proceedings of Participation and Design Conference
  (PDC2002), Sweden}}. \bibinfo{pages}{144--161}.
\newblock
\urldef\tempurl%
\url{http://www.itee.uq.edu.au/~comp4501/_2003/_Readings/GrudinPersonas.pdf}
\showURL{%
\tempurl}


\bibitem[\protect\citeauthoryear{Hargrave and S{\'e}n{\'e}chal}{Hargrave and
  S{\'e}n{\'e}chal}{2000}]%
        {hargrave2000book}
\bibfield{author}{\bibinfo{person}{Anne~C Hargrave} {and}
  \bibinfo{person}{Monique S{\'e}n{\'e}chal}.} \bibinfo{year}{2000}\natexlab{}.
\newblock \showarticletitle{A book reading intervention with preschool children
  who have limited vocabularies: The benefits of regular reading and dialogic
  reading}.
\newblock \bibinfo{journal}{\emph{Early Childhood Research Quarterly}}
  \bibinfo{volume}{15}, \bibinfo{number}{1} (\bibinfo{year}{2000}),
  \bibinfo{pages}{75--90}.
\newblock


\bibitem[\protect\citeauthoryear{Hill, Bordes, Chopra, and Weston}{Hill
  et~al\mbox{.}}{2015}]%
        {hill2015goldilocks}
\bibfield{author}{\bibinfo{person}{Felix Hill}, \bibinfo{person}{Antoine
  Bordes}, \bibinfo{person}{Sumit Chopra}, {and} \bibinfo{person}{Jason
  Weston}.} \bibinfo{year}{2015}\natexlab{}.
\newblock \showarticletitle{The goldilocks principle: Reading children's books
  with explicit memory representations}.
\newblock \bibinfo{journal}{\emph{arXiv preprint arXiv:1511.02301}}
  (\bibinfo{year}{2015}).
\newblock


\bibitem[\protect\citeauthoryear{Hoien and Lundberg}{Hoien and
  Lundberg}{1988}]%
        {hoien1988stages}
\bibfield{author}{\bibinfo{person}{Torleiv Hoien} {and} \bibinfo{person}{Ingvar
  Lundberg}.} \bibinfo{year}{1988}\natexlab{}.
\newblock \showarticletitle{Stages of word recognition in early reading
  development}.
\newblock \bibinfo{journal}{\emph{Scandinavian Journal of Educational
  Research}} \bibinfo{volume}{32}, \bibinfo{number}{4} (\bibinfo{year}{1988}),
  \bibinfo{pages}{163--182}.
\newblock


\bibitem[\protect\citeauthoryear{H\"{o}\"{o}k and L\"{o}wgren}{H\"{o}\"{o}k and
  L\"{o}wgren}{2012}]%
        {hook_2012_strong}
\bibfield{author}{\bibinfo{person}{Kristina H\"{o}\"{o}k} {and}
  \bibinfo{person}{Jonas L\"{o}wgren}.} \bibinfo{year}{2012}\natexlab{}.
\newblock \showarticletitle{Strong Concepts: Intermediate-Level Knowledge in
  Interaction Design Research}.
\newblock \bibinfo{journal}{\emph{ACM Trans. Comput.-Hum. Interact.}}
  \bibinfo{volume}{19}, \bibinfo{number}{3}, Article \bibinfo{articleno}{23}
  (\bibinfo{date}{oct} \bibinfo{year}{2012}), \bibinfo{numpages}{18}~pages.
\newblock
\showISSN{1073-0516}
\urldef\tempurl%
\url{https://doi.org/10.1145/2362364.2362371}
\showDOI{\tempurl}


\bibitem[\protect\citeauthoryear{Horvitz}{Horvitz}{1999}]%
        {horivitz1999principles}
\bibfield{author}{\bibinfo{person}{Eric Horvitz}.}
  \bibinfo{year}{1999}\natexlab{}.
\newblock \showarticletitle{Principles of Mixed-Initiative User Interfaces}. In
  \bibinfo{booktitle}{\emph{Proceedings of the SIGCHI Conference on Human
  Factors in Computing Systems}} (Pittsburgh, Pennsylvania, USA)
  \emph{(\bibinfo{series}{CHI ’99})}. \bibinfo{publisher}{ACM},
  \bibinfo{address}{New York, NY, USA}, \bibinfo{pages}{159–166}.
\newblock
\showISBNx{0201485591}
\urldef\tempurl%
\url{https://doi.org/10.1145/302979.303030}
\showDOI{\tempurl}


\bibitem[\protect\citeauthoryear{Hubbard, Chen, Colunga, Kim, and Yeh}{Hubbard
  et~al\mbox{.}}{2021}]%
        {hubbard2021chid}
\bibfield{author}{\bibinfo{person}{Layne~Jackson Hubbard},
  \bibinfo{person}{Yifan Chen}, \bibinfo{person}{Eliana Colunga},
  \bibinfo{person}{Pilyoung Kim}, {and} \bibinfo{person}{Tom Yeh}.}
  \bibinfo{year}{2021}\natexlab{}.
\newblock \bibinfo{booktitle}{\emph{Child-Robot Interaction to Integrate
  Reflective Storytelling Into Creative Play}}.
\newblock \bibinfo{publisher}{Association for Computing Machinery},
  \bibinfo{address}{New York, NY, USA}.
\newblock
\showISBNx{9781450383769}
\urldef\tempurl%
\url{https://doi.org/10.1145/3450741.3465254}
\showURL{%
\tempurl}


\bibitem[\protect\citeauthoryear{Kim}{Kim}{2017}]%
        {kim2017simple}
\bibfield{author}{\bibinfo{person}{Young-Suk~Grace Kim}.}
  \bibinfo{year}{2017}\natexlab{}.
\newblock \showarticletitle{Why the simple view of reading is not simplistic:
  Unpacking component skills of reading using a direct and indirect effect
  model of reading (DIER)}.
\newblock \bibinfo{journal}{\emph{Scientific Studies of Reading}}
  \bibinfo{volume}{21}, \bibinfo{number}{4} (\bibinfo{year}{2017}),
  \bibinfo{pages}{310--333}.
\newblock


\bibitem[\protect\citeauthoryear{Kory and Breazeal}{Kory and Breazeal}{2014}]%
        {kory2014storytelling}
\bibfield{author}{\bibinfo{person}{Jacqueline Kory} {and}
  \bibinfo{person}{Cynthia Breazeal}.} \bibinfo{year}{2014}\natexlab{}.
\newblock \showarticletitle{Storytelling with robots: Learning companions for
  preschool children's language development}. In \bibinfo{booktitle}{\emph{The
  23rd IEEE international symposium on robot and human interactive
  communication}}. IEEE, \bibinfo{pages}{643--648}.
\newblock


\bibitem[\protect\citeauthoryear{Kotaman}{Kotaman}{2020}]%
        {kotaman2020impacts}
\bibfield{author}{\bibinfo{person}{Huseyin Kotaman}.}
  \bibinfo{year}{2020}\natexlab{}.
\newblock \showarticletitle{Impacts of dialogical storybook reading on young
  children’s reading attitudes and vocabulary development}.
\newblock \bibinfo{journal}{\emph{Reading Improvement}} \bibinfo{volume}{57},
  \bibinfo{number}{1} (\bibinfo{year}{2020}), \bibinfo{pages}{40--45}.
\newblock


\bibitem[\protect\citeauthoryear{Labutov, Basu, and Vanderwende}{Labutov
  et~al\mbox{.}}{2015}]%
        {labutov-etal-2015-deep}
\bibfield{author}{\bibinfo{person}{Igor Labutov}, \bibinfo{person}{Sumit Basu},
  {and} \bibinfo{person}{Lucy Vanderwende}.} \bibinfo{year}{2015}\natexlab{}.
\newblock \showarticletitle{Deep Questions without Deep Understanding}. In
  \bibinfo{booktitle}{\emph{Proceedings of the 53rd Annual Meeting of the
  Association for Computational Linguistics and the 7th International Joint
  Conference on Natural Language Processing (Volume 1: Long Papers)}}.
  \bibinfo{publisher}{ACL}, \bibinfo{address}{Beijing, China},
  \bibinfo{pages}{889--898}.
\newblock
\urldef\tempurl%
\url{https://doi.org/10.3115/v1/P15-1086}
\showDOI{\tempurl}


\bibitem[\protect\citeauthoryear{Lau}{Lau}{2009}]%
        {lau_why_2009}
\bibfield{author}{\bibinfo{person}{Tessa Lau}.}
  \bibinfo{year}{2009}\natexlab{}.
\newblock \showarticletitle{Why {Programming}-{By}-{Demonstration} {Systems}
  {Fail}: {Lessons} {Learned} for {Usable} {AI}}.
\newblock \bibinfo{journal}{\emph{AI Magazine}} \bibinfo{volume}{30},
  \bibinfo{number}{4} (\bibinfo{date}{Oct.} \bibinfo{year}{2009}),
  \bibinfo{pages}{65--67}.
\newblock
\showISSN{0738-4602}
\urldef\tempurl%
\url{http://www.aaai.org/ojs/index.php/aimagazine/article/view/2262}
\showURL{%
\tempurl}


\bibitem[\protect\citeauthoryear{Lazar, Feng, and Hochheiser}{Lazar
  et~al\mbox{.}}{2010}]%
        {Lazar2010ResearchMI}
\bibfield{author}{\bibinfo{person}{Jonathan Lazar}, \bibinfo{person}{Jinjuan
  Feng}, {and} \bibinfo{person}{Harry Hochheiser}.}
  \bibinfo{year}{2010}\natexlab{}.
\newblock \showarticletitle{Research Methods in Human-Computer Interaction}.
\newblock


\bibitem[\protect\citeauthoryear{Learning}{Learning}{2021}]%
        {pillar_learning_meet_2021}
\bibfield{author}{\bibinfo{person}{Pillar Learning}.}
  \bibinfo{year}{2021}\natexlab{}.
\newblock \bibinfo{title}{Meet {Codi} - {An} {Interactive}, {AI}-{Enabled}
  {Smart} {Toy} for {Kids}!}
\newblock
\newblock
\urldef\tempurl%
\url{https://www.pillarlearning.com/}
\showURL{%
\tempurl}


\bibitem[\protect\citeauthoryear{Lever and S{\'e}n{\'e}chal}{Lever and
  S{\'e}n{\'e}chal}{2011}]%
        {lever2011discussing}
\bibfield{author}{\bibinfo{person}{Rosemary Lever} {and}
  \bibinfo{person}{Monique S{\'e}n{\'e}chal}.} \bibinfo{year}{2011}\natexlab{}.
\newblock \showarticletitle{Discussing stories: On how a dialogic reading
  intervention improves kindergartners’ oral narrative construction}.
\newblock \bibinfo{journal}{\emph{Journal of experimental child psychology}}
  \bibinfo{volume}{108}, \bibinfo{number}{1} (\bibinfo{year}{2011}),
  \bibinfo{pages}{1--24}.
\newblock


\bibitem[\protect\citeauthoryear{Lewis, Liu, Goyal, Ghazvininejad, Mohamed,
  Levy, Stoyanov, and Zettlemoyer}{Lewis et~al\mbox{.}}{2019}]%
        {lewis2019bart}
\bibfield{author}{\bibinfo{person}{Mike Lewis}, \bibinfo{person}{Yinhan Liu},
  \bibinfo{person}{Naman Goyal}, \bibinfo{person}{Marjan Ghazvininejad},
  \bibinfo{person}{Abdelrahman Mohamed}, \bibinfo{person}{Omer Levy},
  \bibinfo{person}{Ves Stoyanov}, {and} \bibinfo{person}{Luke Zettlemoyer}.}
  \bibinfo{year}{2019}\natexlab{}.
\newblock \showarticletitle{Bart: Denoising sequence-to-sequence pre-training
  for natural language generation, translation, and comprehension}.
\newblock \bibinfo{journal}{\emph{arXiv preprint arXiv:1910.13461}}
  (\bibinfo{year}{2019}).
\newblock


\bibitem[\protect\citeauthoryear{Li, Azaria, and Myers}{Li
  et~al\mbox{.}}{2017}]%
        {li_sugilite:_2017}
\bibfield{author}{\bibinfo{person}{Toby Jia-Jun Li}, \bibinfo{person}{Amos
  Azaria}, {and} \bibinfo{person}{Brad~A. Myers}.}
  \bibinfo{year}{2017}\natexlab{}.
\newblock \showarticletitle{{SUGILITE}: {Creating} {Multimodal} {Smartphone}
  {Automation} by {Demonstration}}. In \bibinfo{booktitle}{\emph{Proceedings of
  the 2017 {CHI} {Conference} on {Human} {Factors} in {Computing} {Systems}}}
  \emph{(\bibinfo{series}{{CHI} '17})}. \bibinfo{publisher}{ACM},
  \bibinfo{address}{New York, NY, USA}, \bibinfo{pages}{6038--6049}.
\newblock
\showISBNx{978-1-4503-4655-9}
\urldef\tempurl%
\url{https://doi.org/10.1145/3025453.3025483}
\showDOI{\tempurl}


\bibitem[\protect\citeauthoryear{Li, Chen, Xia, Mitchell, and Myers}{Li
  et~al\mbox{.}}{2020a}]%
        {li_sovite:_2020}
\bibfield{author}{\bibinfo{person}{Toby Jia-Jun Li}, \bibinfo{person}{Jingya
  Chen}, \bibinfo{person}{Haijun Xia}, \bibinfo{person}{Tom~M. Mitchell}, {and}
  \bibinfo{person}{Brad~A. Myers}.} \bibinfo{year}{2020}\natexlab{a}.
\newblock \showarticletitle{{Multi-Modal} {Repairs} of {Conversational}
  {Breakdowns} in {Task-Oriented} {Dialogs}}. In
  \bibinfo{booktitle}{\emph{Proceedings of the 33rd {Annual} {ACM} {Symposium}
  on {User} {Interface} {Software} and {Technology}}}
  \emph{(\bibinfo{series}{{UIST} 2020})}. \bibinfo{publisher}{ACM}.
\newblock
\urldef\tempurl%
\url{https://doi.org/10.1145/3379337.3415820}
\showDOI{\tempurl}


\bibitem[\protect\citeauthoryear{Li, Mitchell, and Myers}{Li
  et~al\mbox{.}}{2020b}]%
        {li_interactive:_2020}
\bibfield{author}{\bibinfo{person}{Toby Jia-Jun Li}, \bibinfo{person}{Tom
  Mitchell}, {and} \bibinfo{person}{Brad Myers}.}
  \bibinfo{year}{2020}\natexlab{b}.
\newblock \showarticletitle{Interactive Task Learning from {GUI}-Grounded
  Natural Language Instructions and Demonstrations}. In
  \bibinfo{booktitle}{\emph{Proceedings of the 58th Annual Meeting of the
  Association for Computational Linguistics: System Demonstrations}}.
  \bibinfo{publisher}{ACL}, \bibinfo{address}{Online},
  \bibinfo{pages}{215--223}.
\newblock
\urldef\tempurl%
\url{https://doi.org/10.18653/v1/2020.acl-demos.25}
\showDOI{\tempurl}


\bibitem[\protect\citeauthoryear{Lin, Šabanović, Dombrowski, Miller, Brady,
  and MacDorman}{Lin et~al\mbox{.}}{2021}]%
        {lin_parental:_2021}
\bibfield{author}{\bibinfo{person}{Chaolan Lin}, \bibinfo{person}{Selma
  Šabanović}, \bibinfo{person}{Lynn Dombrowski}, \bibinfo{person}{Andrew~D.
  Miller}, \bibinfo{person}{Erin Brady}, {and} \bibinfo{person}{Karl~F.
  MacDorman}.} \bibinfo{year}{2021}\natexlab{}.
\newblock \showarticletitle{Parental Acceptance of Children’s Storytelling
  Robots: A Projection of the Uncanny Valley of AI}.
\newblock \bibinfo{journal}{\emph{Frontiers in Robotics and AI}}
  \bibinfo{volume}{8} (\bibinfo{year}{2021}), \bibinfo{pages}{49}.
\newblock
\showISSN{2296-9144}
\urldef\tempurl%
\url{https://doi.org/10.3389/frobt.2021.579993}
\showDOI{\tempurl}


\bibitem[\protect\citeauthoryear{Lin}{Lin}{2004}]%
        {lin-2004-rouge}
\bibfield{author}{\bibinfo{person}{Chin-Yew Lin}.}
  \bibinfo{year}{2004}\natexlab{}.
\newblock \showarticletitle{{ROUGE}: A Package for Automatic Evaluation of
  Summaries}. In \bibinfo{booktitle}{\emph{Text Summarization Branches Out}}.
  \bibinfo{publisher}{ACL}, \bibinfo{address}{Barcelona, Spain},
  \bibinfo{pages}{74--81}.
\newblock
\urldef\tempurl%
\url{https://aclanthology.org/W04-1013}
\showURL{%
\tempurl}


\bibitem[\protect\citeauthoryear{Ling}{Ling}{2021}]%
        {ling_luka_2021}
\bibfield{author}{\bibinfo{person}{Ling}.} \bibinfo{year}{2021}\natexlab{}.
\newblock \bibinfo{title}{Luka {AI} reading companion}.
\newblock
\newblock
\urldef\tempurl%
\url{https://luka.ling.ai/}
\showURL{%
\tempurl}


\bibitem[\protect\citeauthoryear{Lovato, Piper, and Wartella}{Lovato
  et~al\mbox{.}}{2019a}]%
        {lovato2019hey}
\bibfield{author}{\bibinfo{person}{Silvia~B Lovato},
  \bibinfo{person}{Anne~Marie Piper}, {and} \bibinfo{person}{Ellen~A
  Wartella}.} \bibinfo{year}{2019}\natexlab{a}.
\newblock \showarticletitle{Hey Google, do unicorns exist? Conversational
  agents as a path to answers to children's questions}. In
  \bibinfo{booktitle}{\emph{Proceedings of the 18th ACM International
  Conference on Interaction Design and Children}}. \bibinfo{pages}{301--313}.
\newblock


\bibitem[\protect\citeauthoryear{Lovato, Piper, and Wartella}{Lovato
  et~al\mbox{.}}{2019b}]%
        {lovato_2019_hey}
\bibfield{author}{\bibinfo{person}{Silvia~B. Lovato},
  \bibinfo{person}{Anne~Marie Piper}, {and} \bibinfo{person}{Ellen~A.
  Wartella}.} \bibinfo{year}{2019}\natexlab{b}.
\newblock \showarticletitle{Hey Google, Do Unicorns Exist? Conversational
  Agents as a Path to Answers to Children's Questions}. In
  \bibinfo{booktitle}{\emph{Proceedings of the 18th ACM International
  Conference on Interaction Design and Children}} (Boise, ID, USA)
  \emph{(\bibinfo{series}{IDC '19})}. \bibinfo{publisher}{Association for
  Computing Machinery}, \bibinfo{address}{New York, NY, USA},
  \bibinfo{pages}{301–313}.
\newblock
\showISBNx{9781450366908}
\urldef\tempurl%
\url{https://doi.org/10.1145/3311927.3323150}
\showDOI{\tempurl}


\bibitem[\protect\citeauthoryear{Luger and Sellen}{Luger and Sellen}{2016}]%
        {Luger2016}
\bibfield{author}{\bibinfo{person}{Ewa Luger} {and} \bibinfo{person}{Abigail
  Sellen}.} \bibinfo{year}{2016}\natexlab{}.
\newblock \showarticletitle{"Like Having a Really Bad PA": The Gulf between
  User Expectation and Experience of Conversational Agents}. In
  \bibinfo{booktitle}{\emph{Proceedings of the 2016 CHI Conference on Human
  Factors in Computing Systems}} (San Jose, California, USA)
  \emph{(\bibinfo{series}{CHI '16})}. \bibinfo{publisher}{Association for
  Computing Machinery}, \bibinfo{address}{New York, NY, USA},
  \bibinfo{pages}{5286–5297}.
\newblock
\showISBNx{9781450333627}
\urldef\tempurl%
\url{https://doi.org/10.1145/2858036.2858288}
\showDOI{\tempurl}


\bibitem[\protect\citeauthoryear{Mack, Rembert, Cummings, and Gilbert}{Mack
  et~al\mbox{.}}{2019}]%
        {mack_codesigning_2019}
\bibfield{author}{\bibinfo{person}{Naja~A. Mack}, \bibinfo{person}{Dekita
  G.~Moon Rembert}, \bibinfo{person}{Robert Cummings}, {and}
  \bibinfo{person}{Juan~E. Gilbert}.} \bibinfo{year}{2019}\natexlab{}.
\newblock \showarticletitle{Co-Designing an Intelligent Conversational History
  Tutor with Children}. In \bibinfo{booktitle}{\emph{Proceedings of the 18th
  ACM International Conference on Interaction Design and Children}} (Boise, ID,
  USA) \emph{(\bibinfo{series}{IDC '19})}. \bibinfo{publisher}{ACM},
  \bibinfo{address}{New York, NY, USA}, \bibinfo{pages}{482–487}.
\newblock
\showISBNx{9781450366908}
\urldef\tempurl%
\url{https://doi.org/10.1145/3311927.3325336}
\showDOI{\tempurl}


\bibitem[\protect\citeauthoryear{Michaelis and Mutlu}{Michaelis and
  Mutlu}{2017}]%
        {michaelis_2017_development}
\bibfield{author}{\bibinfo{person}{Joseph~E. Michaelis} {and}
  \bibinfo{person}{Bilge Mutlu}.} \bibinfo{year}{2017}\natexlab{}.
\newblock \bibinfo{booktitle}{\emph{Someone to Read with: Design of and
  Experiences with an In-Home Learning Companion Robot for Reading}}.
\newblock \bibinfo{publisher}{ACM}, \bibinfo{address}{New York, NY, USA},
  \bibinfo{pages}{301–312}.
\newblock
\showISBNx{9781450346559}
\urldef\tempurl%
\url{https://doi.org/10.1145/3025453.3025499}
\showURL{%
\tempurl}


\bibitem[\protect\citeauthoryear{Mol, Bus, De~Jong, and Smeets}{Mol
  et~al\mbox{.}}{2008}]%
        {mol2008added}
\bibfield{author}{\bibinfo{person}{Suzanne~E Mol}, \bibinfo{person}{Adriana~G
  Bus}, \bibinfo{person}{Maria~T De~Jong}, {and} \bibinfo{person}{Daisy~JH
  Smeets}.} \bibinfo{year}{2008}\natexlab{}.
\newblock \showarticletitle{Added value of dialogic parent--child book
  readings: A meta-analysis}.
\newblock \bibinfo{journal}{\emph{Early education and development}}
  \bibinfo{volume}{19}, \bibinfo{number}{1} (\bibinfo{year}{2008}),
  \bibinfo{pages}{7--26}.
\newblock


\bibitem[\protect\citeauthoryear{Mostow and Chen}{Mostow and Chen}{2009}]%
        {mostow_generationg:_2009}
\bibfield{author}{\bibinfo{person}{Jack Mostow} {and} \bibinfo{person}{Wei
  Chen}.} \bibinfo{year}{2009}\natexlab{}.
\newblock \showarticletitle{Generating Instruction Automatically for the
  Reading Strategy of Self-Questioning}. In
  \bibinfo{booktitle}{\emph{Proceedings of the 2009 Conference on Artificial
  Intelligence in Education: Building Learning Systems That Care: From
  Knowledge Representation to Affective Modelling}}. \bibinfo{publisher}{IOS
  Press}, \bibinfo{address}{NLD}, \bibinfo{pages}{465–472}.
\newblock
\showISBNx{9781607500285}


\bibitem[\protect\citeauthoryear{Muller and Kuhn}{Muller and Kuhn}{1993}]%
        {muller_participatory:_1993}
\bibfield{author}{\bibinfo{person}{Michael~J. Muller} {and}
  \bibinfo{person}{Sarah Kuhn}.} \bibinfo{year}{1993}\natexlab{}.
\newblock \showarticletitle{Participatory Design}.
\newblock \bibinfo{journal}{\emph{Commun. ACM}} \bibinfo{volume}{36},
  \bibinfo{number}{6} (\bibinfo{date}{June} \bibinfo{year}{1993}),
  \bibinfo{pages}{24–28}.
\newblock
\showISSN{0001-0782}
\urldef\tempurl%
\url{https://doi.org/10.1145/153571.255960}
\showDOI{\tempurl}


\bibitem[\protect\citeauthoryear{Myers, Furqan, Nebolsky, Caro, and Zhu}{Myers
  et~al\mbox{.}}{2018}]%
        {myers_2018_patterns}
\bibfield{author}{\bibinfo{person}{Chelsea Myers}, \bibinfo{person}{Anushay
  Furqan}, \bibinfo{person}{Jessica Nebolsky}, \bibinfo{person}{Karina Caro},
  {and} \bibinfo{person}{Jichen Zhu}.} \bibinfo{year}{2018}\natexlab{}.
\newblock \bibinfo{booktitle}{\emph{Patterns for How Users Overcome Obstacles
  in Voice User Interfaces}}.
\newblock \bibinfo{publisher}{Association for Computing Machinery},
  \bibinfo{address}{New York, NY, USA}, \bibinfo{pages}{1–7}.
\newblock
\showISBNx{9781450356206}
\urldef\tempurl%
\url{https://doi.org/10.1145/3173574.3173580}
\showURL{%
\tempurl}


\bibitem[\protect\citeauthoryear{Naderifar, Goli, and Ghaljaie}{Naderifar
  et~al\mbox{.}}{2017}]%
        {Naderifar2017SnowballSA}
\bibfield{author}{\bibinfo{person}{M. Naderifar}, \bibinfo{person}{H. Goli},
  {and} \bibinfo{person}{Fereshteh Ghaljaie}.} \bibinfo{year}{2017}\natexlab{}.
\newblock \showarticletitle{Snowball Sampling: A Purposeful Method of Sampling
  in Qualitative Research}.
\newblock


\bibitem[\protect\citeauthoryear{Newton}{Newton}{1995}]%
        {newton1995role}
\bibfield{author}{\bibinfo{person}{Douglas~P Newton}.}
  \bibinfo{year}{1995}\natexlab{}.
\newblock \showarticletitle{The role of pictures in learning to read}.
\newblock \bibinfo{journal}{\emph{Educational Studies}} \bibinfo{volume}{21},
  \bibinfo{number}{1} (\bibinfo{year}{1995}), \bibinfo{pages}{119--130}.
\newblock


\bibitem[\protect\citeauthoryear{Norman and Malicky}{Norman and
  Malicky}{1987}]%
        {norman1987stages}
\bibfield{author}{\bibinfo{person}{Charles~A Norman} {and}
  \bibinfo{person}{Grace Malicky}.} \bibinfo{year}{1987}\natexlab{}.
\newblock \showarticletitle{Stages in the reading development of adults}.
\newblock \bibinfo{journal}{\emph{Journal of Reading}} \bibinfo{volume}{30},
  \bibinfo{number}{4} (\bibinfo{year}{1987}), \bibinfo{pages}{302--307}.
\newblock


\bibitem[\protect\citeauthoryear{Oviatt}{Oviatt}{1999}]%
        {oviatt_ten_1999}
\bibfield{author}{\bibinfo{person}{Sharon Oviatt}.}
  \bibinfo{year}{1999}\natexlab{}.
\newblock \showarticletitle{Ten {Myths} of {Multimodal} {Interaction}}.
\newblock \bibinfo{journal}{\emph{Commun. ACM}} \bibinfo{volume}{42},
  \bibinfo{number}{11} (\bibinfo{date}{Nov.} \bibinfo{year}{1999}),
  \bibinfo{pages}{74--81}.
\newblock
\showISSN{0001-0782}
\urldef\tempurl%
\url{https://doi.org/10.1145/319382.319398}
\showDOI{\tempurl}


\bibitem[\protect\citeauthoryear{Pantoja, Diederich, Crawford, and
  Hourcade}{Pantoja et~al\mbox{.}}{2019}]%
        {pantoja_voice_2019}
\bibfield{author}{\bibinfo{person}{Luiza~Superti Pantoja},
  \bibinfo{person}{Kyle Diederich}, \bibinfo{person}{Liam Crawford}, {and}
  \bibinfo{person}{Juan~Pablo Hourcade}.} \bibinfo{year}{2019}\natexlab{}.
\newblock \showarticletitle{Voice Agents Supporting High-Quality Social Play}.
  In \bibinfo{booktitle}{\emph{Proceedings of the 18th ACM International
  Conference on Interaction Design and Children}} (Boise, ID, USA)
  \emph{(\bibinfo{series}{IDC '19})}. \bibinfo{publisher}{ACM},
  \bibinfo{address}{New York, NY, USA}, \bibinfo{pages}{314–325}.
\newblock
\showISBNx{9781450366908}
\urldef\tempurl%
\url{https://doi.org/10.1145/3311927.3323151}
\showDOI{\tempurl}


\bibitem[\protect\citeauthoryear{Papineni, Roukos, Ward, and Zhu}{Papineni
  et~al\mbox{.}}{2002}]%
        {papineni_bleu_2002}
\bibfield{author}{\bibinfo{person}{Kishore Papineni}, \bibinfo{person}{Salim
  Roukos}, \bibinfo{person}{Todd Ward}, {and} \bibinfo{person}{Wei-Jing Zhu}.}
  \bibinfo{year}{2002}\natexlab{}.
\newblock \showarticletitle{BLEU: A Method for Automatic Evaluation of Machine
  Translation}. In \bibinfo{booktitle}{\emph{Proceedings of the 40th Annual
  Meeting on Association for Computational Linguistics}} (Philadelphia,
  Pennsylvania) \emph{(\bibinfo{series}{ACL '02})}.
  \bibinfo{publisher}{Association for Computational Linguistics},
  \bibinfo{address}{USA}, \bibinfo{pages}{311–318}.
\newblock
\urldef\tempurl%
\url{https://doi.org/10.3115/1073083.1073135}
\showDOI{\tempurl}


\bibitem[\protect\citeauthoryear{Paris and Paris}{Paris and Paris}{2003}]%
        {paris2003assessing}
\bibfield{author}{\bibinfo{person}{Alison~H Paris} {and}
  \bibinfo{person}{Scott~G Paris}.} \bibinfo{year}{2003}\natexlab{}.
\newblock \showarticletitle{Assessing narrative comprehension in young
  children}.
\newblock \bibinfo{journal}{\emph{Reading Research Quarterly}}
  \bibinfo{volume}{38}, \bibinfo{number}{1} (\bibinfo{year}{2003}),
  \bibinfo{pages}{36--76}.
\newblock


\bibitem[\protect\citeauthoryear{Peck}{Peck}{1989}]%
        {peck1989using}
\bibfield{author}{\bibinfo{person}{Jackie Peck}.}
  \bibinfo{year}{1989}\natexlab{}.
\newblock \showarticletitle{Using storytelling to promote language and literacy
  development}.
\newblock \bibinfo{journal}{\emph{The Reading Teacher}} \bibinfo{volume}{43},
  \bibinfo{number}{2} (\bibinfo{year}{1989}), \bibinfo{pages}{138--141}.
\newblock


\bibitem[\protect\citeauthoryear{Rebanal, Combitsis, Tang, and Chen}{Rebanal
  et~al\mbox{.}}{2021}]%
        {rebanal_2021_xalgo}
\bibfield{author}{\bibinfo{person}{Juan Rebanal}, \bibinfo{person}{Jordan
  Combitsis}, \bibinfo{person}{Yuqi Tang}, {and}
  \bibinfo{person}{Xiang~'Anthony' Chen}.} \bibinfo{year}{2021}\natexlab{}.
\newblock \showarticletitle{XAlgo: A Design Probe of Explaining Algorithms’
  Internal States via Question-Answering}. In \bibinfo{booktitle}{\emph{26th
  International Conference on Intelligent User Interfaces}} (College Station,
  TX, USA) \emph{(\bibinfo{series}{IUI '21})}. \bibinfo{publisher}{Association
  for Computing Machinery}, \bibinfo{address}{New York, NY, USA},
  \bibinfo{pages}{329–339}.
\newblock
\showISBNx{9781450380171}
\urldef\tempurl%
\url{https://doi.org/10.1145/3397481.3450676}
\showDOI{\tempurl}


\bibitem[\protect\citeauthoryear{Rubegni, Dore, Landoni, and Kan}{Rubegni
  et~al\mbox{.}}{2021}]%
        {rubegni2021girl}
\bibfield{author}{\bibinfo{person}{Elisa Rubegni}, \bibinfo{person}{Rebecca
  Dore}, \bibinfo{person}{Monica Landoni}, {and} \bibinfo{person}{Ling Kan}.}
  \bibinfo{year}{2021}\natexlab{}.
\newblock \showarticletitle{“The girl who wants to fly”: Exploring the role
  of digital technology in enhancing dialogic reading}.
\newblock \bibinfo{journal}{\emph{International Journal of Child-Computer
  Interaction}}  \bibinfo{volume}{30} (\bibinfo{year}{2021}),
  \bibinfo{pages}{100239}.
\newblock


\bibitem[\protect\citeauthoryear{Rubegni and Landoni}{Rubegni and
  Landoni}{2014}]%
        {rubegni_2014_fiabot}
\bibfield{author}{\bibinfo{person}{Elisa Rubegni} {and} \bibinfo{person}{Monica
  Landoni}.} \bibinfo{year}{2014}\natexlab{}.
\newblock \showarticletitle{Fiabot! Design and Evaluation of a Mobile
  Storytelling Application for Schools}. In
  \bibinfo{booktitle}{\emph{Proceedings of the 2014 Conference on Interaction
  Design and Children}} (Aarhus, Denmark) \emph{(\bibinfo{series}{IDC '14})}.
  \bibinfo{publisher}{Association for Computing Machinery},
  \bibinfo{address}{New York, NY, USA}, \bibinfo{pages}{165–174}.
\newblock
\showISBNx{9781450322720}
\urldef\tempurl%
\url{https://doi.org/10.1145/2593968.2593979}
\showDOI{\tempurl}


\bibitem[\protect\citeauthoryear{Rubegni, Landoni, Malinverni, and
  Jaccheri}{Rubegni et~al\mbox{.}}{2022}]%
        {RUBEGNI2022102727}
\bibfield{author}{\bibinfo{person}{Elisa Rubegni}, \bibinfo{person}{Monica
  Landoni}, \bibinfo{person}{Laura Malinverni}, {and} \bibinfo{person}{Letizia
  Jaccheri}.} \bibinfo{year}{2022}\natexlab{}.
\newblock \showarticletitle{Raising Awareness of Stereotyping Through
  Collaborative Digital Storytelling: Design for Change with and for Children}.
\newblock \bibinfo{journal}{\emph{International Journal of Human-Computer
  Studies}}  \bibinfo{volume}{157} (\bibinfo{year}{2022}),
  \bibinfo{pages}{102727}.
\newblock
\showISSN{1071-5819}
\urldef\tempurl%
\url{https://doi.org/10.1016/j.ijhcs.2021.102727}
\showDOI{\tempurl}


\bibitem[\protect\citeauthoryear{Scialom, Piwowarski, and Staiano}{Scialom
  et~al\mbox{.}}{2019}]%
        {scialom-etal-2019-self}
\bibfield{author}{\bibinfo{person}{Thomas Scialom}, \bibinfo{person}{Benjamin
  Piwowarski}, {and} \bibinfo{person}{Jacopo Staiano}.}
  \bibinfo{year}{2019}\natexlab{}.
\newblock \showarticletitle{Self-Attention Architectures for Answer-Agnostic
  Neural Question Generation}. In \bibinfo{booktitle}{\emph{Proceedings of the
  57th Annual Meeting of the Association for Computational Linguistics}}.
  \bibinfo{publisher}{ACL}, \bibinfo{address}{Florence, Italy},
  \bibinfo{pages}{6027--6032}.
\newblock
\urldef\tempurl%
\url{https://doi.org/10.18653/v1/P19-1604}
\showDOI{\tempurl}


\bibitem[\protect\citeauthoryear{Shakeri, Nogueira~dos Santos, Zhu, Ng, Nan,
  Wang, Nallapati, and Xiang}{Shakeri et~al\mbox{.}}{2020}]%
        {shakeri-etal-2020-end}
\bibfield{author}{\bibinfo{person}{Siamak Shakeri}, \bibinfo{person}{Cicero
  Nogueira~dos Santos}, \bibinfo{person}{Henghui Zhu}, \bibinfo{person}{Patrick
  Ng}, \bibinfo{person}{Feng Nan}, \bibinfo{person}{Zhiguo Wang},
  \bibinfo{person}{Ramesh Nallapati}, {and} \bibinfo{person}{Bing Xiang}.}
  \bibinfo{year}{2020}\natexlab{}.
\newblock \showarticletitle{End-to-End Synthetic Data Generation for Domain
  Adaptation of Question Answering Systems}. In
  \bibinfo{booktitle}{\emph{Proceedings of the 2020 Conference on Empirical
  Methods in Natural Language Processing (EMNLP)}}. \bibinfo{publisher}{ACL},
  \bibinfo{address}{Online}, \bibinfo{pages}{5445--5460}.
\newblock
\urldef\tempurl%
\url{https://doi.org/10.18653/v1/2020.emnlp-main.439}
\showDOI{\tempurl}


\bibitem[\protect\citeauthoryear{Shanahan and Lonigan}{Shanahan and
  Lonigan}{2010}]%
        {shanahan2010national}
\bibfield{author}{\bibinfo{person}{Timothy Shanahan} {and}
  \bibinfo{person}{Christopher~J Lonigan}.} \bibinfo{year}{2010}\natexlab{}.
\newblock \showarticletitle{The National Early Literacy Panel: A summary of the
  process and the report}.
\newblock \bibinfo{journal}{\emph{Educational Researcher}}
  \bibinfo{volume}{39}, \bibinfo{number}{4} (\bibinfo{year}{2010}),
  \bibinfo{pages}{279--285}.
\newblock


\bibitem[\protect\citeauthoryear{Slov\'{a}k, Theofanopoulou, Cecchet, Cottrell,
  Altarriba~Bertran, Dagan, Childs, and Isbister}{Slov\'{a}k
  et~al\mbox{.}}{2018}]%
        {slovak_child_2018}
\bibfield{author}{\bibinfo{person}{Petr Slov\'{a}k}, \bibinfo{person}{Nikki
  Theofanopoulou}, \bibinfo{person}{Alessia Cecchet}, \bibinfo{person}{Peter
  Cottrell}, \bibinfo{person}{Ferran Altarriba~Bertran}, \bibinfo{person}{Ella
  Dagan}, \bibinfo{person}{Julian Childs}, {and} \bibinfo{person}{Katherine
  Isbister}.} \bibinfo{year}{2018}\natexlab{}.
\newblock \showarticletitle{"I Just Let Him Cry...: Designing Socio-Technical
  Interventions in Families to Prevent Mental Health Disorders}.
\newblock \bibinfo{journal}{\emph{Proc. ACM Hum.-Comput. Interact.}}
  \bibinfo{volume}{2}, \bibinfo{number}{CSCW}, Article \bibinfo{articleno}{160}
  (\bibinfo{date}{nov} \bibinfo{year}{2018}), \bibinfo{numpages}{34}~pages.
\newblock
\urldef\tempurl%
\url{https://doi.org/10.1145/3274429}
\showDOI{\tempurl}


\bibitem[\protect\citeauthoryear{Storch and Whitehurst}{Storch and
  Whitehurst}{2002}]%
        {storch2002oral}
\bibfield{author}{\bibinfo{person}{Stacey~A Storch} {and}
  \bibinfo{person}{Grover~J Whitehurst}.} \bibinfo{year}{2002}\natexlab{}.
\newblock \showarticletitle{Oral language and code-related precursors to
  reading: evidence from a longitudinal structural model.}
\newblock \bibinfo{journal}{\emph{Developmental psychology}}
  \bibinfo{volume}{38}, \bibinfo{number}{6} (\bibinfo{year}{2002}),
  \bibinfo{pages}{934}.
\newblock


\bibitem[\protect\citeauthoryear{Sun and Chai}{Sun and Chai}{2006}]%
        {sun_towards_2006}
\bibfield{author}{\bibinfo{person}{Mingyu Sun} {and} \bibinfo{person}{Joyce~Y.
  Chai}.} \bibinfo{year}{2006}\natexlab{}.
\newblock \showarticletitle{Towards Intelligent QA Interfaces: Discourse
  Processing for Context Questions}. In \bibinfo{booktitle}{\emph{Proceedings
  of the 11th International Conference on Intelligent User Interfaces}}
  (Sydney, Australia) \emph{(\bibinfo{series}{IUI '06})}.
  \bibinfo{publisher}{ACM}, \bibinfo{address}{New York, NY, USA},
  \bibinfo{pages}{163–170}.
\newblock
\showISBNx{1595932879}
\urldef\tempurl%
\url{https://doi.org/10.1145/1111449.1111487}
\showDOI{\tempurl}


\bibitem[\protect\citeauthoryear{Sylla}{Sylla}{2013}]%
        {Sylla2013Design}
\bibfield{author}{\bibinfo{person}{Cristina Sylla}.}
  \bibinfo{year}{2013}\natexlab{}.
\newblock \showarticletitle{Designing a Tangible Interface for Collaborative
  Storytelling to Access 'embodiment' and Meaning Making}. In
  \bibinfo{booktitle}{\emph{Proceedings of the 12th International Conference on
  Interaction Design and Children}} (New York, New York, USA)
  \emph{(\bibinfo{series}{IDC '13})}. \bibinfo{publisher}{Association for
  Computing Machinery}, \bibinfo{address}{New York, NY, USA},
  \bibinfo{pages}{651–654}.
\newblock
\showISBNx{9781450319188}
\urldef\tempurl%
\url{https://doi.org/10.1145/2485760.2485881}
\showDOI{\tempurl}


\bibitem[\protect\citeauthoryear{Tamura, Kimoto, Shiomi, Iio, Shimohara, and
  Hagita}{Tamura et~al\mbox{.}}{2017}]%
        {tamura2017effects}
\bibfield{author}{\bibinfo{person}{Yumiko Tamura}, \bibinfo{person}{Mitsuhiko
  Kimoto}, \bibinfo{person}{Masahiro Shiomi}, \bibinfo{person}{Takamasa Iio},
  \bibinfo{person}{Katsunori Shimohara}, {and} \bibinfo{person}{Norihiro
  Hagita}.} \bibinfo{year}{2017}\natexlab{}.
\newblock \showarticletitle{Effects of a Listener Robot with Children in
  Storytelling}. In \bibinfo{booktitle}{\emph{Proceedings of the 5th
  International Conference on Human Agent Interaction}} (Bielefeld, Germany)
  \emph{(\bibinfo{series}{HAI '17})}. \bibinfo{publisher}{ACM},
  \bibinfo{address}{New York, NY, USA}, \bibinfo{pages}{35–43}.
\newblock
\showISBNx{9781450351133}
\urldef\tempurl%
\url{https://doi.org/10.1145/3125739.3125750}
\showDOI{\tempurl}


\bibitem[\protect\citeauthoryear{Tang, Duan, Qin, Yan, and Zhou}{Tang
  et~al\mbox{.}}{2017}]%
        {tang2017question}
\bibfield{author}{\bibinfo{person}{Duyu Tang}, \bibinfo{person}{Nan Duan},
  \bibinfo{person}{Tao Qin}, \bibinfo{person}{Zhao Yan}, {and}
  \bibinfo{person}{Ming Zhou}.} \bibinfo{year}{2017}\natexlab{}.
\newblock \showarticletitle{Question answering and question generation as dual
  tasks}.
\newblock \bibinfo{journal}{\emph{arXiv preprint arXiv:1706.02027}}
  (\bibinfo{year}{2017}).
\newblock


\bibitem[\protect\citeauthoryear{Vasalou, Kalantari, Kucirkova, and
  Vezzoli}{Vasalou et~al\mbox{.}}{2020}]%
        {vasalou2020designing}
\bibfield{author}{\bibinfo{person}{Asimina Vasalou}, \bibinfo{person}{Sara
  Kalantari}, \bibinfo{person}{Natalia Kucirkova}, {and}
  \bibinfo{person}{Yvonne Vezzoli}.} \bibinfo{year}{2020}\natexlab{}.
\newblock \showarticletitle{Designing for oral storytelling practices at home:
  A parental perspective}.
\newblock \bibinfo{journal}{\emph{International Journal of Child-Computer
  Interaction}}  \bibinfo{volume}{26} (\bibinfo{year}{2020}),
  \bibinfo{pages}{100214}.
\newblock


\bibitem[\protect\citeauthoryear{Wang, Churchill, Maes, Fan, Shneiderman, Shi,
  and Wang}{Wang et~al\mbox{.}}{2020}]%
        {wang_human_2020}
\bibfield{author}{\bibinfo{person}{Dakuo Wang}, \bibinfo{person}{Elizabeth
  Churchill}, \bibinfo{person}{Pattie Maes}, \bibinfo{person}{Xiangmin Fan},
  \bibinfo{person}{Ben Shneiderman}, \bibinfo{person}{Yuanchun Shi}, {and}
  \bibinfo{person}{Qianying Wang}.} \bibinfo{year}{2020}\natexlab{}.
\newblock \showarticletitle{From Human-Human Collaboration to Human-AI
  Collaboration: Designing AI Systems That Can Work Together with People}. In
  \bibinfo{booktitle}{\emph{Extended Abstracts of the 2020 CHI Conference on
  Human Factors in Computing Systems}} (Honolulu, HI, USA)
  \emph{(\bibinfo{series}{CHI EA '20})}. \bibinfo{publisher}{Association for
  Computing Machinery}, \bibinfo{address}{New York, NY, USA},
  \bibinfo{pages}{1–6}.
\newblock
\showISBNx{9781450368193}
\urldef\tempurl%
\url{https://doi.org/10.1145/3334480.3381069}
\showDOI{\tempurl}


\bibitem[\protect\citeauthoryear{Wang, Maes, Ren, Shneiderman, Shi, and
  Wang}{Wang et~al\mbox{.}}{2021}]%
        {wang_designing_2021}
\bibfield{author}{\bibinfo{person}{Dakuo Wang}, \bibinfo{person}{Pattie Maes},
  \bibinfo{person}{Xiangshi Ren}, \bibinfo{person}{Ben Shneiderman},
  \bibinfo{person}{Yuanchun Shi}, {and} \bibinfo{person}{Qianying Wang}.}
  \bibinfo{year}{2021}\natexlab{}.
\newblock \bibinfo{booktitle}{\emph{Designing AI to Work WITH or FOR People?}}
\newblock \bibinfo{publisher}{ACM}, \bibinfo{address}{New York, NY, USA}.
\newblock
\showISBNx{9781450380959}
\urldef\tempurl%
\url{https://doi.org/10.1145/3411763.3450394}
\showURL{%
\tempurl}


\bibitem[\protect\citeauthoryear{Wang, Yuan, and Trischler}{Wang
  et~al\mbox{.}}{2017}]%
        {wang2017joint}
\bibfield{author}{\bibinfo{person}{Tong Wang}, \bibinfo{person}{Xingdi Yuan},
  {and} \bibinfo{person}{Adam Trischler}.} \bibinfo{year}{2017}\natexlab{}.
\newblock \showarticletitle{A joint model for question answering and question
  generation}.
\newblock \bibinfo{journal}{\emph{arXiv preprint arXiv:1706.01450}}
  (\bibinfo{year}{2017}).
\newblock


\bibitem[\protect\citeauthoryear{Wiederhold}{Wiederhold}{2018}]%
        {wiederhold2018alexa}
\bibfield{author}{\bibinfo{person}{Brenda~K Wiederhold}.}
  \bibinfo{year}{2018}\natexlab{}.
\newblock \bibinfo{title}{“Alexa, are you my Mom?” The role of artificial
  intelligence in child development}.
\newblock
\newblock


\bibitem[\protect\citeauthoryear{Wright}{Wright}{1995}]%
        {wright1995storytelling}
\bibfield{author}{\bibinfo{person}{Andrew Wright}.}
  \bibinfo{year}{1995}\natexlab{}.
\newblock \bibinfo{booktitle}{\emph{Storytelling with children}}.
\newblock \bibinfo{publisher}{Oxford University}.
\newblock


\bibitem[\protect\citeauthoryear{Xu}{Xu}{2020}]%
        {xu2020using}
\bibfield{author}{\bibinfo{person}{Ying Xu}.} \bibinfo{year}{2020}\natexlab{}.
\newblock \showarticletitle{Using conversational agents to foster young
  children's science learning from screen media}. In
  \bibinfo{booktitle}{\emph{Proceedings of the 2020 ACM Interaction Design and
  Children Conference: Extended Abstracts}}. \bibinfo{pages}{14--19}.
\newblock


\bibitem[\protect\citeauthoryear{Xu, Branham, Deng, Collins, and Warschauer}{Xu
  et~al\mbox{.}}{2021a}]%
        {xu_current:_2021}
\bibfield{author}{\bibinfo{person}{Ying Xu}, \bibinfo{person}{Stacy Branham},
  \bibinfo{person}{Xinwei Deng}, \bibinfo{person}{Penelope Collins}, {and}
  \bibinfo{person}{Mark Warschauer}.} \bibinfo{year}{2021}\natexlab{a}.
\newblock \bibinfo{booktitle}{\emph{Are Current Voice Interfaces Designed to
  Support Children’s Language Development?}}
\newblock \bibinfo{publisher}{Association for Computing Machinery},
  \bibinfo{address}{New York, NY, USA}.
\newblock
\showISBNx{9781450380966}
\urldef\tempurl%
\url{https://doi.org/10.1145/3411764.3445271}
\showURL{%
\tempurl}


\bibitem[\protect\citeauthoryear{Xu, Wang, Collins, Lee, and Warschauer}{Xu
  et~al\mbox{.}}{2021b}]%
        {xu_same:_2021}
\bibfield{author}{\bibinfo{person}{Ying Xu}, \bibinfo{person}{Dakuo Wang},
  \bibinfo{person}{Penelope Collins}, \bibinfo{person}{Hyelim Lee}, {and}
  \bibinfo{person}{Mark Warschauer}.} \bibinfo{year}{2021}\natexlab{b}.
\newblock \showarticletitle{Same benefits, different communication patterns:
  Comparing Children's reading with a conversational agent vs. a human
  partner}.
\newblock \bibinfo{journal}{\emph{Computers \& Education}}
  \bibinfo{volume}{161} (\bibinfo{year}{2021}), \bibinfo{pages}{104059}.
\newblock
\showISSN{0360-1315}
\urldef\tempurl%
\url{https://doi.org/10.1016/j.compedu.2020.104059}
\showDOI{\tempurl}


\bibitem[\protect\citeauthoryear{Xu and Warschauer}{Xu and Warschauer}{2019}]%
        {xu_2019_young}
\bibfield{author}{\bibinfo{person}{Ying Xu} {and} \bibinfo{person}{Mark
  Warschauer}.} \bibinfo{year}{2019}\natexlab{}.
\newblock \showarticletitle{Young Children's Reading and Learning with
  Conversational Agents}. In \bibinfo{booktitle}{\emph{Extended Abstracts of
  the 2019 CHI Conference on Human Factors in Computing Systems}} (Glasgow,
  Scotland Uk) \emph{(\bibinfo{series}{CHI EA '19})}.
  \bibinfo{publisher}{Association for Computing Machinery},
  \bibinfo{address}{New York, NY, USA}, \bibinfo{pages}{1–8}.
\newblock
\showISBNx{9781450359719}
\urldef\tempurl%
\url{https://doi.org/10.1145/3290607.3299035}
\showDOI{\tempurl}


\bibitem[\protect\citeauthoryear{Xu and Warschauer}{Xu and Warschauer}{2020a}]%
        {xu2020exploring}
\bibfield{author}{\bibinfo{person}{Ying Xu} {and} \bibinfo{person}{Mark
  Warschauer}.} \bibinfo{year}{2020}\natexlab{a}.
\newblock \showarticletitle{Exploring young children's engagement in joint
  reading with a conversational agent}. In
  \bibinfo{booktitle}{\emph{Proceedings of the Interaction Design and Children
  Conference}}. \bibinfo{pages}{216--228}.
\newblock


\bibitem[\protect\citeauthoryear{Xu and Warschauer}{Xu and Warschauer}{2020b}]%
        {xu_exploring_2020}
\bibfield{author}{\bibinfo{person}{Ying Xu} {and} \bibinfo{person}{Mark
  Warschauer}.} \bibinfo{year}{2020}\natexlab{b}.
\newblock \showarticletitle{Exploring Young Children's Engagement in Joint
  Reading with a Conversational Agent}. In
  \bibinfo{booktitle}{\emph{Proceedings of the Interaction Design and Children
  Conference}} (London, United Kingdom) \emph{(\bibinfo{series}{IDC '20})}.
  \bibinfo{publisher}{Association for Computing Machinery},
  \bibinfo{address}{New York, NY, USA}, \bibinfo{pages}{216–228}.
\newblock
\showISBNx{9781450379816}
\urldef\tempurl%
\url{https://doi.org/10.1145/3392063.3394417}
\showDOI{\tempurl}


\bibitem[\protect\citeauthoryear{Xu and Warschauer}{Xu and Warschauer}{2020c}]%
        {xu2020you}
\bibfield{author}{\bibinfo{person}{Ying Xu} {and} \bibinfo{person}{Mark
  Warschauer}.} \bibinfo{year}{2020}\natexlab{c}.
\newblock \showarticletitle{What Are You Talking To?: Understanding Children's
  Perceptions of Conversational Agents}. In
  \bibinfo{booktitle}{\emph{Proceedings of the 2020 CHI Conference on Human
  Factors in Computing Systems}}. \bibinfo{pages}{1--13}.
\newblock


\bibitem[\protect\citeauthoryear{Yao, Wang, Wu, Hoang, Sun, Li, Yu, and Xu}{Yao
  et~al\mbox{.}}{2021}]%
        {yao2021ais}
\bibfield{author}{\bibinfo{person}{Bingsheng Yao}, \bibinfo{person}{Dakuo
  Wang}, \bibinfo{person}{Tongshuang Wu}, \bibinfo{person}{Tran Hoang},
  \bibinfo{person}{Branda Sun}, \bibinfo{person}{Toby Jia-Jun Li},
  \bibinfo{person}{Mo Yu}, {and} \bibinfo{person}{Ying Xu}.}
  \bibinfo{year}{2021}\natexlab{}.
\newblock \bibinfo{title}{It is AI's Turn to Ask Human a Question: Question and
  Answer Pair Generation for Children Storybooks in FairytaleQA Dataset}.
\newblock
\newblock
\showeprint[arxiv]{2109.03423}~[cs.CL]


\bibitem[\protect\citeauthoryear{Yao, Bouma, and Zhang}{Yao
  et~al\mbox{.}}{2012}]%
        {yao2012semantics}
\bibfield{author}{\bibinfo{person}{Xuchen Yao}, \bibinfo{person}{Gosse Bouma},
  {and} \bibinfo{person}{Yi Zhang}.} \bibinfo{year}{2012}\natexlab{}.
\newblock \showarticletitle{Semantics-based question generation and
  implementation}.
\newblock \bibinfo{journal}{\emph{Dialogue \& Discourse}} \bibinfo{volume}{3},
  \bibinfo{number}{2} (\bibinfo{year}{2012}), \bibinfo{pages}{11--42}.
\newblock


\bibitem[\protect\citeauthoryear{Zevenbergen and Whitehurst}{Zevenbergen and
  Whitehurst}{2003}]%
        {zevenbergen2003dialogic}
\bibfield{author}{\bibinfo{person}{Andrea~A Zevenbergen} {and}
  \bibinfo{person}{Grover~J Whitehurst}.} \bibinfo{year}{2003}\natexlab{}.
\newblock \showarticletitle{Dialogic reading: A shared picture book reading
  intervention for preschoolers}.
\newblock \bibinfo{journal}{\emph{On reading books to children: Parents and
  teachers}} (\bibinfo{year}{2003}), \bibinfo{pages}{177--200}.
\newblock


\end{thebibliography}

\end{document}